\documentclass[12pt,a4j]{article}
\usepackage[dvips]{graphicx}
\usepackage{enumerate}
\usepackage{amsmath}
\usepackage{amssymb}
\usepackage{amsfonts}

\begin{document}
\baselineskip=7mm
\centerline{\bf Smooth multisoliton solutions and their peakon limit} \par
\centerline{\bf of Novikov's Camassa-Holm type equation with cubic nonlinearity }\par

\bigskip
\centerline{Yoshimasa Matsuno\footnote{{\it E-mail address}: matsuno@yamaguchi-u.ac.jp}}\par

\centerline{\it Division of Applied Mathematical Science,}\par
\centerline{\it Graduate School of Science and Engineering} \par
\centerline{\it Yamaguchi University, Ube, Yamaguchi 755-8611, Japan} \par
\bigskip

\leftline{\bf Abstract}\par
We consider Novikov's Camassa-Holm type equation with cubic nonlinearity. In particular, we present
a compact parametric representation of the smooth bright multisolution solutions on a constant background  and
investigate their structure. We find that the tau-functions associated with the solutions are closely
related to those of a model equation for  shallow-water waves (SWW) introduced by Hirota and Satsuma.
This novel feature is established by applying the reciprocal transformation to the Novikov equation.
We also show  by specifying a complex phase parameter that the  smooth soliton is converted to a novel  singular soliton with single cusp and double peaks.
We   demonstrate that both the smooth and singular solitons converge to a peakon as the background field tends to zero
whereby we employ a method that has been developed for performing the similar limiting procedure for the multisoliton solutions of 
the Camassa-Holm equation. 
In the subsequent asymptotic analysis of the two- and $N$-soliton solutions, we confirm their solitonic behaviour.
Remarkably, the formulas for the phase shifts of solitons as well as their peakon limits  coincide formally with those of the Degasperis-Procesi equation.
Last, we derive an infinite number of conservation laws of the Novikov equation by using a relation between
solutions of  the Novikov equation and those of the SWW equation.
In appendix, we prove various bilinear identities associated with the tau-functions of the multisoliton solutions of the SWW equation.

\newpage

\leftline{\bf 1. Introduction} \par
\bigskip
\noindent Recently, Novikov introduced an integrable Camassa-Holm (CH) type equation (the Novikov equation hereafter)   in an attempt to classify the nonlocal 
partial differential equations (PDEs) with quadratic or cubic nonlinearity [1]. 
It may be written in the form
$$m_t+u^2m_x+3uu_xm=0,\qquad  m=u-u_{xx}, \eqno(1.1)$$
where $u=u(x,t)$ is a  function of  time $t$ and a spatial variable $x$, and the subscripts $x$ and $t$ appended to $m$ and $u$ denote partial differentiation.
 Using the perturbative symmetry approach, he found a few symmetries and then derived a scalar Lax representation for it. 
 Subsequently, Hone and Wang gave a matrix Lax representation and showed that the Novikov equation is related by a reciprocal transformation to
 a negative flow in an integrable hierarchy of the Sawada-Kotera equation [2]. A bi-Hamiltonian structure of the hierarchy was also provided as well as 
 an infinite number of conservation laws. A remarkable feature of the Novikov equation is that it admits peaked waves (or peakons) whose dynamics are shown to
 obey a finite dimensional integrable Hamiltonian system [2, 3]. As for exact solutions,  a few works have been concerned with peakons [2-4]. 
 We emphasize that  smooth soliton solutions have not been available as yet. \par
 The purpose of this paper is to construct the  smooth $N$-soliton solution ($N$: arbitrary positive integer) of the Novikov equation which satisfies the boundary condition 
$u \rightarrow u_0$\ ($u_0$: real constant) as $|x| \rightarrow \infty$ and analyze its structure. 
Note that equation (1.1) is invariant under the transformation $u\rightarrow -u$. Hence, we may consider positive solutions only. The constant $u_0$ is therefore taken to be positive
without loss of generality.
While the peakon solutions of the Novikov equation  have been constructed under the vanishing boundary condition,
we demonstrate for the first time that the smooth solitons  converge to the peakons in the
limit of zero background field ($u_0\rightarrow 0$). \par
This paper is organized as follows. In section 2, we transform equation (1.1) to a system of equations by a reciprocal transformation and reveal that it can be solved in terms of
the tau-function associated with
 the $N$-soliton solution of a model equation for shallow-water waves (SWW equation for short)  introduced by Hirota and Satsuma [5]. 
The various novel bilinear identities are presented among the tau-functions which are related simply to the tau-function for the $N$-soliton solution of the SWW equation.
In section 3, we provide a compact parametric representation for the $N$-soliton solution of the Novikov equation. 
The proof of the solution is performed by means of a purely algebraic procedure within the framework of the bilinear formalism.
We find that the structure of the tau-functions
for the $N$-soliton solution is relevant to that of the Degasperis-Procesi (DP) equation [6, 7].
In section 4, we investigate the properties of the one-, two- and $N$-soliton
 solutions  in detail. The smooth one-soliton solution takes the form of a bright soliton on a constant background. 
  In addition, we show that a novel  singular soliton with W-shaped profile is produced from the smooth soliton which
  exhibits both cusp and peak. This can be established simply by specifying a complex phase parameter for the smooth soliton solution.
Last, we demonstrate that the smooth soliton  converges to the peakon in the limit of  $u_0\rightarrow 0$  with the velocity of the
soliton being fixed, which we call the peakon limit. This limiting procedure is performed by employing
a novel method [8, 9] that has been developed for solving the similar problem for the multisoliton solutions of 
the CH equation. We also provide a numerical evidence for the passage to the peakon.
Furthermore, we briefly discuss the passage of the singular soliton to a peakon in the same context.
In the subsequent asymptotic analysis of the
two- and $N$-soliton solutions, we obtain the formulas for the phase shift of  solitons and confirm their solitonic behaviour. 
Remarkably, we notice that the formulas coincide formally with those of the
$N$-soliton solution of the DP equation [6, 7]. These formulas reduce, in the peakon limit, to the corresponding ones for the two- and $N$-peakon solutions of the Novikov equation obtained  by means of the
inverse scattering transform (IST) method [3].
In section 5, we derive an infinite number of conservation laws starting from those of the SWW equation. While the conservation laws have been constructed by using the Lax pair for the Novikov equation [2], 
our method is based on  a purely algebraic procedure without recourse to the  IST.
Section 6 is devoted to concluding remarks. In appendix, we prove various bilinear identities presented in section 2.
\par
\bigskip
\leftline{\bf 2. Reciprocal transformation and SWW equation} \par
\medskip
\leftline{\it 2.1.  Reciprocal transformation}\par
\bigskip
\noindent In accordance with [2], we introduce  the coordinate transformation $(x, t)\rightarrow (y, \tau)$ by
$$dy=m^{2/3}\,dx-m^{2/3}u^2\,dt,\qquad d\tau=dt, \eqno(2.1a)$$
subjected to the restriction $m>0$.
Consequently, the $x$ and $t$ derivatives can be rewritten as $${\partial\over\partial x}
=m^{2/3}\,{\partial\over\partial y},\qquad {\partial\over\partial t}=\,{\partial\over\partial \tau}-m^{2/3}u^2\,{\partial\over\partial y}. \eqno(2.1b)$$
It now follows from (2.1b) that the variable $x=x(y,\tau)$ satisfies a  system of linear PDEs
$$x_y=m^{-2/3},\qquad x_\tau=u^2. \eqno(2.2)$$
We apply the transformation (2.1b) to the Novikov equation and find that it can be recast into the form
$$m_\tau+3m^{5/3}uu_y=0. \eqno(2.3)$$
On the other hand, $u$ from (1.1) can be rewritten in terms of $m$ as
$$u=m+m^{4/3}u_{yy}+{2\over 3}m^{1/3}m_yu_y. \eqno(2.4)$$
If we define the new variables $V$ and $W$ by $V=m^{2/3}$ and $W=um^{1/3}$, respectively, then equations (2.3) and (2.4)
can be put into the form [2]
$$\left({1\over V}\right)_\tau=\left({W^2\over V}\right)_y, \eqno(2.5)$$
$$W_{yy}+UW+1=0, \eqno(2.6a)$$
where
$$U=-{V_{yy}\over 2V}+{V_y^2\over 4V^2}-
{1\over V^2}. \eqno(2.6b)$$
\par
The following proposition comes from the compatibility condition $\psi_{\tau yyy}=\psi_{yyy\tau}$ of the Lax pair for  the Novikov equation written in terms of the variables $y$ and $\tau$ [2]
$$\psi_{yyy}+U\psi_y=\lambda^2\psi,\qquad \psi_\tau={1\over \lambda^2}\,(W\psi_{yy}-W_y\psi_y)-{2\over 3\lambda^2}\,\psi. \eqno(2.7)$$
Here, we provide an alternative proof based on (2.5) and (2.6). \par
\bigskip
\noindent {\bf Proposition 2.1.}\ {\it The variables $U$ and $W$ satisfies a linear PDE}  
$$U_\tau+3W_y=0. \eqno(2.8) $$
\bigskip
\noindent{\bf Proof.} First, we put $p=1/V, Z=W^2$ and rewrite (2.5) and (2.6b) in terms of $p$ and $Z$ as
$$p_\tau=(pZ)_y, \qquad U={p_{yy}\over 2p}-{3\over 4}{p_y^2\over p^2}-p^2. \eqno(2.9)$$
A direct computation using the above two equations, as well as equation (2.6a) to replace the term $W_{yy}$ yields
$$U_\tau={1\over 2}Z_{yyy}+2UZ_y+U_yZ,\qquad Z_{yyy}=-4UZ_y-2U_yZ-6W_y. \eqno(2.10)$$
Both expressions of (2.10) immediately lead to equation (2.8).     \hspace{\fill}$\Box$ \par
\bigskip
If we  eliminate the variable $W$ from (2.6a) and (2.8), we obtain a single equation for $U$:
$$UU_{\tau yy}-U_yU_{\tau y}+U^2U_\tau+3U_y=0. \eqno(2.11)$$
Note that the evolution equation for $p$ in the independent variables $\tau$ and $y$ is the reciprocal transformation of the 
Novikov equation which can be derived by substituting $U$ from (2.9) into equation (2.11).
  The resultant expression is, however, too formidable to write down explicitly. \par
\bigskip
\leftline{\it 2.2. SWW equation}\par
\medskip
\noindent We show that the system of equations (2.6a) and (2.8) for $U$ and $W$
 admits the $N$-soliton solution by reducing it to the SWW equation. 
To this end, we first seek the $N$-soliton solution of equation (2.11) of the form
$$U=U_0+6\,({\rm ln}\,f)_{yy}, \qquad f=f(y, \tau). \eqno(2.12)$$
The above dependent variable transformation
enables us to recast (2.11) to the bilinear equation for $f$
$$(D_\tau D_y^3-3W_0D_y^2+U_0D_\tau D_y)f\cdot f=0. \eqno(2.13)$$
Here, the bilinear operators $D_y$ and $D_\tau$ are defined by
$$D_y^mD_\tau^nf\cdot g=\left({\partial\over \partial y}-{\partial\over \partial y^\prime}\right)^m\left({\partial\over \partial\tau}-{\partial\over \partial\tau^\prime}\right)^n
f(y, \tau)g(y^\prime, \tau^\prime)\Big|_{y^\prime=y, \tau^\prime=\tau}, \eqno(2.14)$$
where $m$ and $n$ are nonnegative integers.
The constants $U_0$ and $W_0$ are boundary values of $U$ and $W$, respectively as $|y| \rightarrow \infty$. Specifically, since $u\rightarrow u_0$ and $m\rightarrow u_0$ in this limit,
it follows from the definition of $V(=m^{2/3})$, $W(=um^{1/3})$ and $U$ from (2.6b) that $U_0=-u_0^{-4/3}$ and $W_0=u_0^{4/3}$. 
The tau-function $f$ introduced in (2.12) will be shown to be the most important ingredient in constructing the $N$-soliton solution of the Novikov equation. \par
A remarkable feature of  equation (2.13) is that it coincides with the bilinear form of the SWW equation introduced by Hirota and Satsuma [5]. Actually,
 by means of the dependent variable transformation $q=2({\rm ln}\,f)_{yy}$, as well as the formulas 
 $${D_\tau D_yf\cdot f\over f^2}=2\,({\rm ln}\,f)_{\tau y},\qquad { D_y^2f\cdot f\over f^2}=2\,({\rm ln}\,f)_{yy}, \eqno(2.15a)$$
 $${D_\tau D_y^3f\cdot f\over f^2}=2\,({\rm ln}\,f)_{\tau yyy}+12\,({\rm ln}\,f)_{\tau y}({\rm ln}\,f)_{yy}, \eqno(2.15b)$$
 the bilinear equation (2.13) can be transformed,
after rewriting it in terms of the variable $q$, to the SWW equation
$$q_\tau+3\kappa^4\,q_y-3\kappa^2qq_\tau+3\kappa^2\,q_y\int_y^\infty q_\tau dy-\kappa^2q_{\tau yy}=0, \qquad q=q(y, \tau), \eqno(2.16)$$
where the positive parameter $\kappa$ has been introduced for later convenience by the relation $\kappa=u_0^{2/3}$ so that $U_0=-\kappa^{-2}$ and $W_0=\kappa^2$.
Substituting (2.12) into equation (2.8) and integrating once with respect to $y$ under the boundary condition $W\rightarrow \kappa^2, |y|\rightarrow \infty$, 
we obtain the expression of $W$ in terms of the tau-function $f$
$$W=\kappa^2-2({\rm ln}\,f)_{\tau y}. \eqno(2.17)$$
\par
Last, it follows from (2.2) and the definition of $W$ that the variable $x=x(y, \tau)$ obeys the  linear PDE
$$x_\tau=W^2x_y. \eqno(2.18)$$
If one can solve equation (2.18) for given $W$, then the expression of $u$ follows immediately from the second equation of (2.2). This provides
a parametric representation for  the $N$-soliton solution of the Novikov equation. 
Recall that the integrability of equation (2.18) (or equivalently, that of the system of equations (2.2)) is assured by (2.3). The method of solution for it is the core in the present analysis.
\par
\medskip
\noindent{\bf Remark 2.1.}\ It is important that $U$ and $q$ are connected by
$$U=-\kappa^{-2}+3q, \eqno(2.19) $$
which follows from (2.12) and the transformation $q=2({\rm ln}\,f)_{yy}$. The relation (2.19) can be interpreted as a B\"acklund transformation between solutions of the
equation for $p(=1/V)$ and those of the SWW equation via $U$ from (2.9). It can be used to derive an infinite number of conservation laws of the Novikov equation, as will be demonstrated in section 5. \par
\bigskip
\leftline{\it2.3.  Bilinear identities for the tau-functions}\par
\noindent The tau-function $f$
 for the $N$-solton solution of the SWW equation  is given compactly by [5]
$$f=\sum_{\mu=0,1}{\rm exp}\left[\sum_{i=1}^N\mu_i\xi_i+\sum_{1\leq i<j\leq N}\mu_i\mu_j\gamma_{ij}\right], \eqno(2.20a)$$
with
$$\xi_i=k_i\left[y-{3\kappa^4\over 1-(\kappa k_i)^2}\tau-y_{i0}\right],\qquad (i=1, 2, ..., N),\eqno(2.20b)$$
$${\rm e}^{\gamma_{ij}}={(k_i-k_j)^2[(k_i^2-k_ik_j+k_j^2)\kappa^2-3]\over (k_i+k_j)^2[(k_i^2+k_ik_j+k_j^2)\kappa^2-3]}, \qquad (i, j=1, 2, ..., N; i\not=j).\eqno(2.20c)$$
Here, $k_i$ and $y_{i0}$ are the amplitude and phase parameters of the $i$th soliton, respectively,
and the notation $\sum_{\mu=0, 1}$ implies the summation over all possible combination of $\mu_1=0, 1, \mu_2=0, 1, ..., \mu_N=0, 1$. 
We shall prove in appendix that the tau-function $f$ solves the bilinear equation (2.13). \par
To proceed, let us introduce some notations. The $N$-soliton  solution from (2.20) is parametrized by
the $N$ phase variables $\xi_i\ (i=1, 2, ..., N)$ and hence we use  a vector notation $f=f({\mathbf\xi})$ with an $N$-component  row vector ${\mathbf\xi}=(\xi_1, \xi_2, ..., \xi_N)$.
Let ${\mathbf\phi}=(\phi_1, \phi_2, ..., \phi_N)$ be  an $N$-component  row vector with the elements
$${\rm e}^{-\phi_i}=\sqrt{{(1-{\kappa k_i\over 2})(1-\kappa k_i)\over (1+{\kappa k_i\over 2})(1+\kappa k_i)}},\qquad (i=1, 2, ..., N).\eqno(2.21)$$
Define the tau-functions $f_1, f_1^\prime, f_2$ and $f_2^\prime$ by making use of the above notation
$$f_1=f({\mathbf\xi}-{\mathbf\phi}), \quad f_1^\prime=f({\mathbf\xi}-2{\mathbf\phi}),\quad f_2=f({\mathbf\xi}+{\mathbf\phi}),\quad f_2^\prime=f({\mathbf\xi}+2{\mathbf\phi}). \eqno(2.22)$$
Then, the following proposition provides various identities among the tau-functions (2.22).
. \par
\noindent {\bf Proposition 2.2.}\ {\it The tau-functions $f, f_1^\prime$ and $f_2^\prime$ satisfy the bilinear identities}
$$D_yf_1^\prime\cdot f_2^\prime+{2\over\kappa}f_1^\prime f_2^\prime={2\over\kappa^3}(\kappa^2 f^2-D_\tau D_yf\cdot f), \eqno(2.23)$$
$$D_\tau f_1^\prime\cdot f_2^\prime+2\kappa^3f_1^\prime f_2^\prime={2\over\kappa^3}(\kappa^6 f^2+D_\tau^2f\cdot f), \eqno(2.24)$$
$$D_y^3 f_1^\prime\cdot f_2^\prime+{6\over\kappa}D_y^2f_1^\prime\cdot f_2^\prime+{11\over\kappa^2}D_yf_1^\prime\cdot f_2^\prime+{6\over\kappa^3}(f_1^\prime f_2^\prime-f^2)=0, \eqno(2.25)$$
$$D_\tau f_1^\prime\cdot f_2^\prime+\kappa D_\tau D_y f_1^\prime\cdot f_2^\prime+{\kappa^2\over 4}D_\tau D_y^2f_1^\prime\cdot f_2^\prime+\kappa^3(f_1^\prime f_2^\prime-f^2)
+{\kappa^4\over 2}D_yf_1^\prime\cdot f_2^\prime+{\kappa^5\over 2}D_y^2f_1^\prime\cdot f_2^\prime$$
$$={1\over 2\kappa}(D_\tau^2D_y^2f\cdot f+\kappa^6D_y^2f\cdot f). \eqno(2.26)$$
The proof of the above proposition will be presented in appendix. These bilinear identities as well as (2.13) play the central role in constructing the $N$-soliton solution
of the Novikov equation. \par
\bigskip
\leftline{\bf 3. The $N$-soliton solution}\par
\bigskip
\noindent Let us introduce the tau-function $g=g({\mathbf \xi})$
$$g=\sum_{\mu, \nu=0, 1}{\rm exp}\Biggl[\sum_{i=1}^N(\mu_i+\nu_i)\xi_i+\sum_{i=1}^N(2\mu_i\nu_i-\mu_i-\nu_i)\,{\rm ln}\,a_i$$
$$+{1\over 2}\sum_{\substack{ i, j=1\\ (i\not=j)}}^N(\mu_i\mu_j+\nu_i\nu_j)A_{2i-1, 2j-1}+{1\over 2}\sum_{\substack{ i, j=1\\ (i\not=j)}}^N(\mu_i\nu_j+\mu_j\nu_i)A_{2i-1, 2j}\Biggr]. \eqno(3.1a)$$
Here
$$a_i=\sqrt{{1-{\kappa^2k_i^2\over 4}\over 1-\kappa^2k_i^2}},\qquad (i= 1, 2, ..., N), \eqno(3.1b)$$
$${\rm exp}\left[A_{2i-1, 2j-1}\right]={(p_i-p_j)(q_i-q_j)\over (p_i+q_j)(q_i+p_j)}, \qquad (i, j=1, 2, ..., N; i\not=j).\eqno(3.1c)$$
$${\rm exp}\left[A_{2i-1, 2j}\right]={(p_i-q_j)(q_i-p_j)\over (p_i+q_j)(q_i+q_j)}, \qquad (i, j=1, 2, ..., N; i\not=j).\eqno(3.1d)$$
$$p_i={k_i\over 2}\left[1+{2\over \kappa k_i}\sqrt{{1\over 3}\left(1-{1\over 4}\kappa^2 k_i^2\right)}\right],\qquad (i= 1, 2, ..., N), \eqno(3.1e)$$
$$q_i={k_i\over 2}\left[1-{2\over \kappa k_i}\sqrt{{1\over 3}\left(1-{1\over 4}\kappa^2 k_i^2\right)}\right],\qquad (i= 1, 2, ..., N), \eqno(3.1f)$$
and $\xi_i\ (i=1, 2, ..., N)$ are already given by (2.20b). \par
The tau-functions $g_1$ and $g_2$ are defined by
$$g_1=g({\mathbf \xi}-{\mathbf \phi}), \qquad g_2=g({\mathbf \xi}+{\mathbf \phi}), \eqno(3.2)$$
where ${\mathbf \phi}$ is the $N$-component row vector  introduced by (2.21). \par
\medskip
\noindent {\bf Remark 3.1.}\ The tau-function $g$ has already appeared in constructing the $N$-soliton solution of the DP equation. See  the expression (2.11) of [7],
where the notation $f$ is used in place of $g$. \par
\medskip
Now,  the main result in our paper is given by the following theorem.
\par
\bigskip
\noindent{\bf Theorem 3.1.}\ {\it The Novikov equation (1.1) admits the parametric representation for the $N$-soliton solution
$$u^2=u^2(y, \tau)=\kappa^3+{1\over 2}{\partial\over \partial \tau}\,{\rm ln}\,{g_1\over g_2}, \eqno(3.3a)$$
$$x=x(y, \tau)={y\over\kappa}+\kappa^3 \tau+{1\over 2}\,{\rm ln}\,{g_1\over g_2}+d, \eqno(3.3b)$$
where the tau-fuctions $g_1$ and $g_2$ are given by (3.1) and (3.2) and $d$ is an arbitrary constant.} \par
\bigskip
The proof of theorem 3.1 will be carried out by a sequence of steps. We shall start our discussion with the proposition 3.1.   \par
\bigskip
\noindent {\bf Proposition 3.1.}\ {\it The following relation holds among the tau-functions $g$, $f_1$ and $f_2$ 
$$g=f_1f_2+\kappa D_yf_1\cdot f_2, \eqno(3.4)$$
where $f_1$ and $f_2$ are defined by (2.22).} \par
\bigskip
\noindent {\bf Proof.}\ This  relation stems from  (3.13) of [7] if one replaces $f$, $g_1$ $g_2$ by $g$, $f_1$ and $f_2$, respectively. \hspace{\fill}$\Box$ \par
\bigskip
If we use (3.4) and take into account the notation (2.22), we can express $g_1=g(\xi-\phi)$ and $g_2=g(\xi+\phi)$ in terms of $f$, $f_1^\prime$ and $f_2^\prime $. Explicitly,
$$g_1=f_1^\prime f+\kappa D_yf_1^\prime \cdot f, \eqno(3.5a)$$
$$g_2=f_2^\prime f-\kappa D_yf_2^\prime \cdot f. \eqno(3.5b)$$
\bigskip
The proposition below connects the tau-functions $g_1$ and $g_2$ with the tau-function $f$. \par
\medskip
\noindent {\bf Proposition 3.2.}\ {\it The tau-functions $f$, $g_1$ and $g_2$ satisfy the relations}
$$\left(D_y+{2\over\kappa}\right)g_1\cdot g_2={2\over\kappa}f^4, \eqno(3.6a)$$
$$\left(D_\tau+2\kappa^3\right)g_1\cdot g_2={2\over\kappa}(\kappa^2f^2-D_\tau D_yf\cdot f)^2. \eqno(3.6b)$$
\bigskip
\noindent{\bf Proof.}\ First, we prove (3.6a). Substituting (3.5) into (3.6a), we obtain, after some straightforward calculations, 
$$\left(D_y+{2\over\kappa}\right)g_1\cdot g_2-{2\over\kappa}f^4$$
$$={\kappa^2\over 4}\left[D_y^3 f_1^\prime\cdot f_2^\prime+{6\over\kappa}D_y^2f_1^\prime\cdot f_2^\prime+{11\over\kappa^2}D_yf_1^\prime\cdot f_2^\prime+{6\over\kappa^3}(f_1^\prime f_2^\prime-f^2)\right]f^2 $$
$$+\kappa^2\left(-{1\over 4}fF_{yy}+f_yF_y-f_{yy}F+{1\over 4\kappa^2}\,fF-{1\over 2\kappa^3}\,f^3\right)f, \eqno(3.7)$$
where we have put $F=D_yf_1^\prime \cdot f_2^\prime+(2/\kappa)f_1^\prime f_2^\prime$ for simplicity. Since $F$ is equal to $(2/\kappa^3)(\kappa^2 f^2-D_\tau D_yf\cdot f)$ by (2.23), the second term on the right-hand side of (3.7)
divided by $\kappa^2 f$ simplifies to
$$-{1\over 4}fF_{yy}+f_yF_y-f_{yy}F+{1\over 4\kappa^2}\,fF-{1\over 2\kappa^3}\,f^3=-{1\over 2\kappa^3}(W_{yy}+UW+1)f^3,$$
after using (2.12) and (2.17). This expression becomes zero by (2.6a).  The first term of (3.7) also turns out to be zero by virtue of (2.25), completing the proof of (3.6a). \par
To prove (3.6b), let $P=\left(D_\tau+2\kappa^3\right)g_1\cdot g_2-{2\over\kappa}(\kappa^2f^2-D_\tau D_yf\cdot f)^2$. This expression can be modified, after introducing (3.5) into it, to
$P=P_1+P_2$ with
$$P_1=\kappa^2(-f_y^2G+ff_yG_y+{1\over 2}FD_\tau D_yf\cdot f),$$
$$P_2=\left[D_\tau f_1^\prime\cdot f_2^\prime+\kappa D_\tau D_y f_1^\prime\cdot f_2^\prime+\kappa^2(-f_{1,\tau y}^\prime f_{2,y}^\prime+f_{1,y}^\prime f_{2,\tau y}^\prime)+\kappa^4D_y f_1^\prime\cdot f_2^\prime
-2\kappa^5 f_{1,y}^\prime f_{2,y}^\prime\right]f^2,$$
where $G=D_\tau f_1^\prime\cdot f_2^\prime+2\kappa^3f_1^\prime f_2^\prime$.  Using the right-hand side of (2.23) for $F$  and that of (2.24) for $G$, respectively, $P_1$ becomes
$$P_1={2\over\kappa}\left[2f_yf_{\tau \tau y}-2f_{\tau y}^2+{\kappa^4\over 2}\left(f^2-f_1^\prime f_2^\prime-{\kappa\over 2}D_yf_1^\prime\cdot f_2^\prime\right)+\kappa^6f_y^2\right]f^2. $$
Let $\bar P=P/f^2$. Then,
$$\bar P={2\over\kappa}(2f_yf_{\tau \tau y}-2f_{\tau y}^2+\kappa^6f_y^2)$$
$$+D_\tau f_1^\prime\cdot f_2^\prime+\kappa D_\tau D_y f_1^\prime\cdot f_2^\prime+\kappa^2(-f_{1,\tau y}^\prime f_{2,y}^\prime+f_{1,y}^\prime f_{2,\tau y}^\prime)
+\kappa^3(f^2-f_1^\prime f_2^\prime)+{\kappa^4\over 2}D_y f_1^\prime\cdot f_2^\prime
-2\kappa^5f_{1,y}^\prime f_{2,y}^\prime.$$
Differentiating (2.24) twice with respect to $y$, we deduce
$$\kappa^2(-f_{1,\tau y}^\prime f_{2,y}^\prime+f_{1,y}^\prime f_{2,\tau y}^\prime)={\kappa^2\over 4}D_\tau D_y^2f_1^\prime\cdot f_2^\prime+{\kappa^5\over 2}(f_1^\prime f_2^\prime)_{yy}
-{1\over 2\kappa}(\kappa^6 f^2+D_\tau^2f\cdot f)_{yy},$$
which, substituted into the corresponding term in $\bar P$, gives
$$\bar P=-{1\over 2\kappa}(D_\tau^2D_y^2f\cdot f+\kappa^6D_y^2f\cdot f)$$
$$+D_\tau f_1^\prime\cdot f_2^\prime+\kappa D_\tau D_y f_1^\prime\cdot f_2^\prime+{\kappa^2\over 4}D_\tau D_y^2f_1^\prime\cdot f_2^\prime+\kappa^3(f^2-f_1^\prime f_2^\prime)
+{\kappa^4\over 2}D_yf_1^\prime\cdot f_2^\prime+{\kappa^5\over 2}D_y^2f_1^\prime\cdot f_2^\prime.$$
This expression vanishes  by virtue of (2.26), completing the proof of (3.6b). \hspace{\fill}$\Box$ \par
\bigskip
We are now ready for proving theorem 3.1. Actually, it is a consequence of proposition 3.2. \par
\bigskip
\noindent{\bf Proof of Theorem 3.1.}\ We establish the theorem by showing that  expression (3.3b) for $x$ satisfies equation (2.18). 
To this end, we rewrite the latter equation as $u^2=W^2x_y$ by referring to the second equation of (2.2).
If we substitute (2.17), (3.3a) and (3.3b) into this equation, then the equation to be proved becomes
$$(D_\tau +2\kappa^3)g_1\cdot g_2={1\over f^4}(\kappa^2f^2-D_\tau D_yf\cdot f)^2\left(D_y+{2\over\kappa}\right)g_1\cdot g_2.$$
In view of (3.6a) and (3.6b), the above equation holds identically. This completes the proof of theorem 3.1. \hspace{\fill}$\Box$ \par
\bigskip
\noindent {\bf Remark 3.2.}\ The determinant of the transformation matrix corresponding to the coordinate transformation (2.1a) is $m^{2/3}(=1/x_y)$. 
With use of (3.3b) and (3.6a), this can be evaluated as $\kappa g_1g_2/f^4$, which would turn out to be a positive quantity
provided that the conditions $0<\kappa k_i<1\ (i=1, 2, ..., N)$ are satisfied for the amplitude parameters of solitons. In this setting, the mapping (2.1a) becomes one-to-one. 
Then, the expression of $u$ from (3.3) gives rise to a single-valued function of $x$. Explicit examples will be presented in the  next section for the one- and two-soliton solutions. \par
\bigskip
\leftline{\bf 4. Properties of soliton solutions}\par
\medskip
\leftline{\it 4.1. One-soliton solution }\par
\leftline{\it 4.1.1. Smooth soliton} \par
\noindent The tau-functions corresponding to the one-soliton solution are given from (3.1) and (3.2) (or from (2.20), (2.22) and (3.5) ) with $N=1$. They read
$$g_1=1+{4(1-\alpha)\over 2+\alpha}\,{\rm e}^\xi+{2-\alpha\over 2+\alpha}{1-\alpha\over 1+\alpha}\,{\rm e}^{2\xi}, \eqno(4.1a)$$
$$g_2=1+{4(1+\alpha)\over 2-\alpha}\,{\rm e}^\xi+{2+\alpha\over 2-\alpha}{1+\alpha\over 1-\alpha}\,{\rm e}^{2\xi}, \eqno(4.1b)$$
with
$$\xi=k(y-\tilde c\tau-y_0),\qquad \tilde c={3\kappa^4\over 1-\alpha^2}, \eqno(4.1c)$$
where we have put $\xi=\xi_1, k=k_1, \alpha=\kappa k_1$ and $y_0=y_{10}$ for simplicity. We assume $k>0$ hereafter
and the condition $0<\alpha<1$ is imposed to assure the smoothness of the solution. \par
The parametric representation of the smooth one-soliton solution follows from (3.3) and (4.1). It can be written in the form

\begin{align*}
u^2  &=\kappa^3+{12k\alpha \tilde c\over 4-\alpha^2}{\cosh\,\xi+{1\over 2}{2+\alpha^2\over 1-\alpha^2}
\over \cosh\,2\xi+{8(2+\alpha^2)\over 4-\alpha^2}\,\cosh\,\xi
+{3(4-\alpha^2+3\alpha^4)\over(1-\alpha^2)(4-\alpha^2)}} \\
        &= {2\kappa^3\left(\cosh\,\xi+{1+2\alpha^2\over 1-\alpha^2}\right)^2 
        \over \cosh\,2\xi+{8(2+\alpha^2)\over 4-\alpha^2}\,\cosh\,\xi
+{3(4-\alpha^2+3\alpha^4)\over(1-\alpha^2)(4-\alpha^2)}}, \tag*{(4.2a)}
\end{align*}

$$X\equiv x-ct-x_0={\xi\over\alpha}+{1\over 2}\,{\rm ln}\left({\tanh^2{\xi\over 2}-{2\over\alpha}\tanh\,{\xi\over2}+{4-\alpha^2\over 3\alpha^2}
\over \tanh^2{\xi\over 2}+{2\over\alpha}\tanh\,{\xi\over2}+{4-\alpha^2\over 3\alpha^2}}\right), \eqno(4.2b)$$
where
$$c={\tilde c\over \kappa}+\kappa^3={\kappa^3(4-\alpha^2)\over 1-\alpha^2}, \eqno(4.2c)$$
is the vecolity of the soliton in the $(x, t)$ coordinate system and $x_0=y_0/\kappa$.
The constant $d$ has been  chosen such that $\xi=0$ corresponds to $X=0$. 
Notice that the form of $u^2$ in the second line of (4.2a) can be anticipated from (3.3a) and (3.6b). Actually, the numerator of $u^2$
is represented by a square of exponential functions.
\par

\begin{figure}[t]
\begin{center}
\includegraphics[width=10cm]{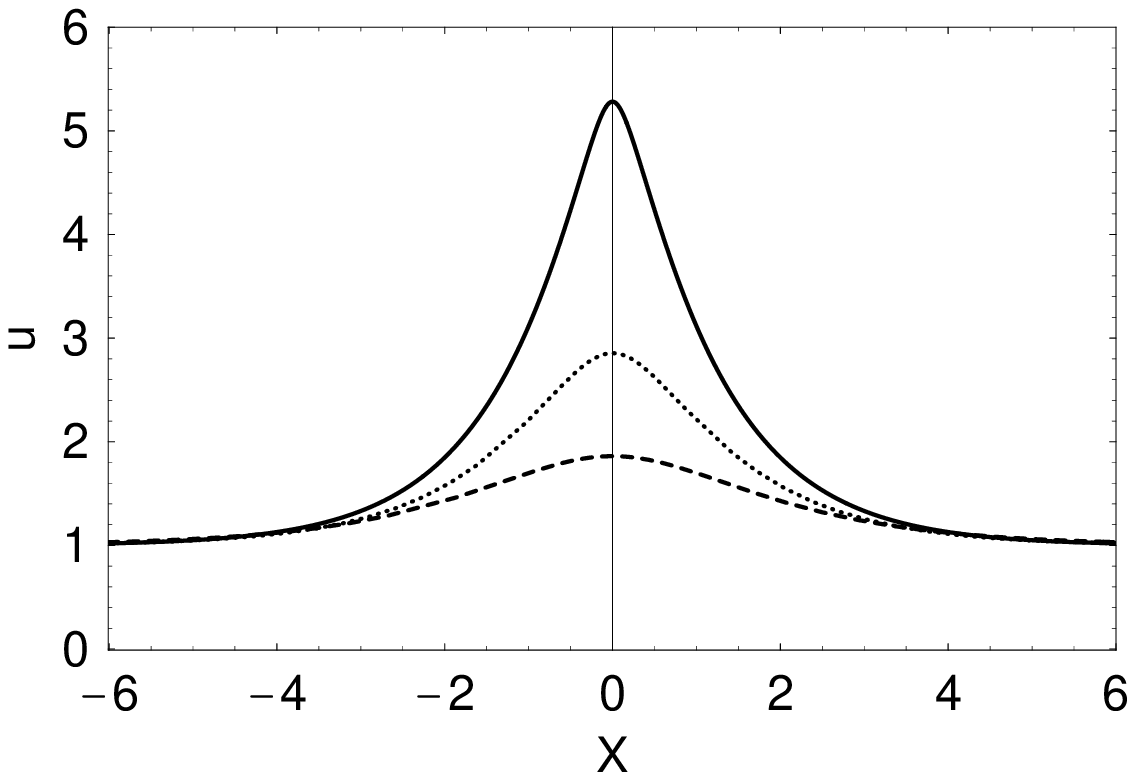}
\end{center}
\noindent {{\bf Figure 1.}\ The profile of  smooth solitons with $\kappa=1$. $\alpha=0.7$\,(dashed curve), $\alpha=0.85$\,(dotted curve), $\alpha=0.95$\,(solid curve).}
\end{figure}

Figure 1 depicts the profile of  smooth solitons against the stationary coordinate $X$ for three distinct values of $\alpha$ with $\kappa=1$.
The one-soliton solution represents a bright soliton on a constant  background $u=\kappa^{3/2}$ whose center position $x_c$ is located at
$x_c=ct+x_0$.  The amplitude of the soliton with respective to the background field, which we denote by $A$, is
found to be as
$$A=\kappa^{3/2}\left({2+\alpha^2\over \sqrt{(1-\alpha^2)(4-\alpha^2)}}-1\right). \eqno(4.3)$$
Eliminating the parameter $\alpha$ from (4.2c) and (4.3), we obtain the amplitude-velocity relation
$$c={1\over 2}\left[(A+\kappa^{3/2})^2+4\kappa^3+(A+\kappa^{3/2})\sqrt{(A+\kappa^{3/2})^2+8\kappa^3}\right]. \eqno(4.4)$$
We see from this expression that the velocity becomes a monotonically increasing function of the amplitude. 
Only in the small amplitude limit, (4.4) recovers a linear relation between $c$ and $A$.
Of particular interest is the limit $\kappa\rightarrow 0$ for which $c$ is equal to $A^2$.
 This limiting value
of the velocity coincides with the velocity of the peakon, as will be revealed in section 4.1.3. \par
 To investigate the feature of the solution in more detail, we particularly focus on the profile of $u$ near the crest.  This can be accomplished if one expands $u$ and $X$ 
with respect to $\xi$ as
$$u=c^{1/2}\left[{2+\alpha^2\over 4-\alpha^2}-{9\alpha^2(1-\alpha^2)\over (4-\alpha^2)^3}\,\xi^2+O(\xi^4)\right],\eqno(4.5a)$$
$$X={4(1-\alpha^2)\over \alpha(4-\alpha^2)}\,\xi+{2\alpha(2-\alpha^2-\alpha^4)\over (4-\alpha^2)^3}\,\xi^3+O(\xi^5). \eqno(4.5b)$$
By eliminating the variable $\xi$ from (4.5a) and (4.5b), we can see that near the crest $X\sim 0$, $u$ and its $X$ derivative  behave like
$$u=c^{1/2}\left[{2+\alpha^2\over 4-\alpha^2}-{9\over 16}\,{\alpha^4\over (1-\alpha^2)(4-\alpha^2)}\,X^2+O(X^4)\right],\eqno(4.6a)$$
$${du\over dX}=-{9\over 8}{\alpha^4c^{1/2}\over (1-\alpha^2)(4-\alpha^2)}\,X+O(X^3).\eqno(4.6b)$$
The expression (4.6b) indicates that as $\alpha$ increases, the crest of the smooth soliton becomes sharp. 
Note, however that the expansion breaks down in the vicinity of $\alpha=1$ for which we need a separate treatment.
The asymptotic behaviour of $u$ as $\alpha$ tends to 1 will be
explored in detail  in section 4.1.3, showing that it forms a peak. \par
\bigskip
\leftline{\it 4.1.2. Singular soliton}\par
\noindent The singular soliton is obtained from the smooth soliton (4.2) if one replaces the phase variable $x_0$ and  $y_0$ by $x_0+\pi i/\alpha$ and $y_0+\pi i/k$, respectively.
 In this setting, $\cosh\,\xi\rightarrow -\cosh\,\xi$ and $\tanh(\xi/2)\rightarrow \coth(\xi/2)$,
 giving rise to the parametric representation of $u^2$
 \begin{align*}
u^2 &=\kappa^3+{12k\alpha \tilde c\over 4-\alpha^2}{-\cosh\,\xi+{1\over 2}{2+\alpha^2\over 1-\alpha^2}
\over \cosh\,2\xi-{8(2+\alpha^2)\over 4-\alpha^2}\,\cosh\,\xi
+{3(4-\alpha^2+3\alpha^4)\over(1-\alpha^2)(4-\alpha^2)}} \\
 &= {2\kappa^3\left(-\cosh\,\xi+{1+2\alpha^2\over 1-\alpha^2}\right)^2 
        \over \cosh\,2\xi-{8(2+\alpha^2)\over 4-\alpha^2}\,\cosh\,\xi
+{3(4-\alpha^2+3\alpha^4)\over(1-\alpha^2)(4-\alpha^2)}}, \tag*{(4.7a)}
\end{align*}
$$X\equiv x-ct-x_0={\xi\over\alpha}+{1\over 2}\,{\rm ln}\left({\coth^2{\xi\over 2}-{2\over\alpha}\coth\,{\xi\over2}+{4-\alpha^2\over 3\alpha^2}
\over \coth^2{\xi\over 2}+{2\over\alpha}\coth\,{\xi\over2}+{4-\alpha^2\over 3\alpha^2}}\right). \eqno(4.7b)$$
\par
Figure 2 shows the typical profile of  singular solitons for three distinct values of $\alpha$ with $\kappa=1$. 
We can observe that the  singularities appear both at the crest $X=0$ and at $X=\pm X_0$, where $X_0$ is a positive
constant specified later.\par
\begin{figure}[t]
\begin{center}
\includegraphics[width=10cm]{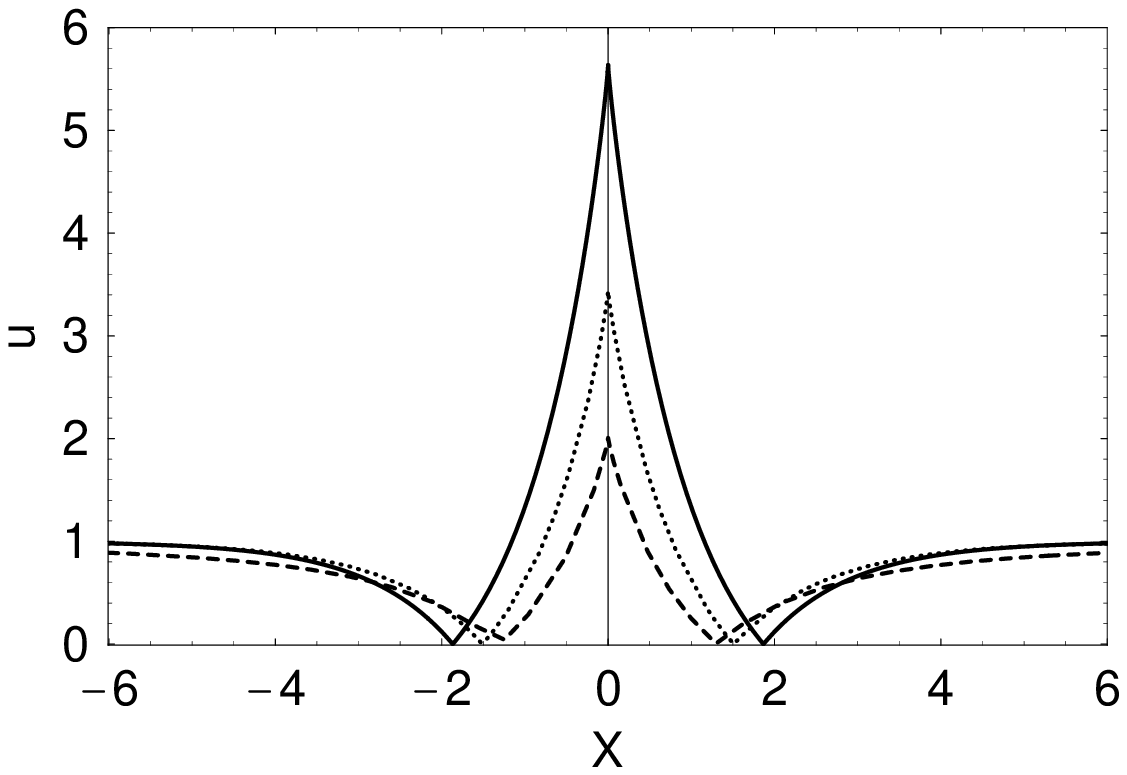}
\end{center}
\noindent {{\bf Figure 2.}\ The profile of  singular solitons with $\kappa=1$. $\alpha=0.1$\,(dashed curve), $\alpha=0.85$\,(dotted curve), $\alpha=0.95$\,(solid curve).}
\end{figure}
 We first analyze the structure of the soliton near the crest. Expanding $u$ and $X$ near $\xi=0$, we obtain
$$u=c^{1/2}\left[1-{1-\alpha^2\over 24 \alpha^4}\,\xi^4+O(\xi^6)\right], \eqno(4.8a)$$
$$X={(1-\alpha^2)(4-\alpha^2)\over 180 \alpha^5}\,\xi^5+{(1-\alpha^2)(2-\alpha^2)(4-\alpha^2)\over 1512 \alpha^7}\,\xi^7+O(\xi^9). \eqno(4.8b)$$
Elimination of the variable $\xi$ from the above expressions yields the approximate expressions of $u$ and $du/dX$ near $X=0$. They read
$$u=c^{1/2}\left[1-{180^{4/5}\over 24}\,{(1-\alpha^2)^{1/5}\over (4-\alpha^2)^{4/5}}\,X^{4/5}+O(X^{6/5})\right],\eqno(4.9a)$$
$${du\over dX}=-{180^{4/5}\over 30}\,{(1-\alpha^2)^{1/5}c^{1/2}\over (4-\alpha^2)^{4/5}}\,X^{-1/5}+O(X^{1/5}). \eqno(4.9b)$$
We find from (4.9b) that
 $du/dX\rightarrow \mp\infty$ as $X\rightarrow\pm 0$, showing 
 that the crest of the soliton takes the form of a cusp. \par
To investigate the structure of the soliton near $X=\pm X_0$, we first observe that $u$ from (4.7a) has two zeros $\xi=\pm\xi_0$ with
$\xi_0=\cosh^{-1}[(1+2\alpha^2)/(1-\alpha^2)]$. The corresponding value of $X_0$ is given from (4.7b) by
$$X_0={1\over\alpha}\,{\rm ln}\left[{1+2\alpha^2+\alpha\sqrt{3(2+\alpha^2)}\over 1-\alpha^2}\right]
+{1\over 2}\,{\rm ln}\left[{3-\sqrt{3(2+\alpha^2)}\over 3+\sqrt{3(2+\alpha^2)}}\right]. \eqno(4.10)$$
We expand $u$ and $X$ near $\xi=\xi_0$ to obtain
$$u=c^{1/2}\left[{\sqrt{(1-\alpha^2)(2+\alpha^2)}\over 3\alpha^2}|\xi-\xi_0|+{1\over 6\sqrt{3}}(4+\alpha^2-2\alpha^4)\sqrt{1-\alpha^2}\,(\xi-\xi_0)^2+O\big((\xi-\xi_0)^2\big)\right],\eqno(4.11a)$$
$$X=X_0+{1+3\alpha-\alpha^2\over 3\alpha^2}\,(\xi-\xi_0)+O\big((\xi-\xi_0)^2\big).\eqno(4.11b)$$
It turns out from (4.11) that
$$u=c^{1/2}\left[{\sqrt{(1-\alpha^2)(2+\alpha^2)}\over 1+3\alpha-\alpha^2}|X-X_0|+O\big((X-X_0)^2\big)\right],\eqno(4.12a)$$
$${du\over dX}=c^{1/2}\left[{\sqrt{(1-\alpha^2)(2+\alpha^2)}\over 1+3\alpha-\alpha^2}\,{\rm sgn}(X-X_0)+O(X-X_0))\right], \eqno(4.12b)$$
where ${\rm sign}\, X$ denotes the sign function, i.e., ${\rm sign}\, X=1$ for $X>0$ and ${\rm sign}\, X=-1$ for $X<0$.
The corresponding expressions of $u, X$ and $du/dX$ near $\xi=-\xi_0$ or $X=-X_0$ follows from (4.11) and (4.12) if one replaces $\xi_0$ and $X_0$ by $-\xi_0$ and $-X_0$, respectively. 
The expression (4.12a) indicates that $u$ exhibits a peak at $X=\pm X_0$. Notice from (4.10) that $X_0$ is a monotonically increasing function of $\alpha$ and has 
a limiting value 1.303 when $\alpha\rightarrow 0$.
The profile of $u$ with $\alpha=0.1$ drawn in figure 2 is closest to the limiting form as $\alpha\rightarrow 0$.
\par
\bigskip
\leftline{\it 4.1.3. Peakon}\par
\noindent It has been shown that the Novikov equation admits no smooth solutions  which vanish at infinity. Under the same boundary condition, however it exhibits
a peaked wave (or peakon) solution of the form [2-4]
$$u=\sqrt{c}\,{\rm e}^{-|x-ct-x_0|}. \eqno(4.13)$$
More generally, the Novikov equation has the multipeakon solutions whose dynamics are governed by an integrable finite dimensional dynamical system for the positions  and  amplitudes of peakons [2, 3].
Here, we demonstrate that the the single peakon solution can be reduced from the smooth one-soliton solution by taking an appropriate limit.  The similar limiting procedure has been performed
for the smooth multisoliton solutions of the CH equation [8, 9]. Here, we apply the method developed in [9] to obtain the peakon solution (4.13). \par
First, we observe that the tau-function $g_2$ from (4.1b) exhibits a singularity at $\alpha=1$. 
This becomes an obstacle in the limiting process. Hence,
we remove it by
 replacing the phase variable as $\xi\rightarrow \xi-\phi$, where
${\rm e}^{-\phi}=\sqrt{[(1-\alpha/2)(1-\alpha)/(1+\alpha/2)(1+\alpha)]}$ (see (2.21)). Consequently, the tau-functions $g_1$ and $g_2$ can be put into the form
$$g_1=1+2\left({1-\alpha\over 1+{\alpha\over 2}}\right)^{3/2}\left({1-{\alpha\over 2}\over  1+\alpha}\right)^{1/2}\,{\rm e}^\xi
+\left[{(1-{\alpha\over 2})(1-\alpha)\over (1+{\alpha\over 2})(1+\alpha)}\right]^2\,e^{2\xi}, \eqno(4.14a)$$
$$g_2=1+2\left({1-\alpha^2\over 1-{\alpha^2\over 4}}\right)^{1/2}\,{\rm e}^\xi+\,e^{2\xi}. \eqno(4.14b)$$
\par
We now take the limit $\kappa\rightarrow 0$ with $c$ being fixed, where
$c$ is the velocity of the smooth soliton in the $(x, t)$ coordinate system which is given by (4.2c.). The constancy of $c$   demands  that
one must takes the limit  $\alpha\rightarrow 1$ simultaneously. By eliminating the variable $y$ from (3.3b) and (4.1c) and subsitutiong the result for $\xi$ into ${\rm e}^\xi$, we obtain
$${\rm e}^\xi = \left(g_2\over g_1\right)^{\alpha/2}\,z^\alpha, \qquad z={\rm e}^{x-ct-x_0},\qquad \xi=\alpha\left(x-ct-{1\over 2}\,{\rm ln}{g_1\over g_2}-x_0\right), \qquad x_0=d+{y_0\over\kappa}. \eqno(4.15)$$
Note from (4.14) that both $g_1$ and $g_2$ are positive and hence $g_2/g_1$ is a positive quantity.
In the limit $\kappa\rightarrow 0$, $\alpha$ from (4.2c) is expanded in powers of $\kappa$ as $\alpha=1-(3/2c)\,\kappa^3+O(\kappa^6)$. 
Substituting this expansion and (4.15) into (4.14), the tau-functions $g_1$ and $g_2$  are  approximated by
$$g_1\sim 1+{\kappa^{9/2}\over c^{3/2}}\sqrt{g_2\over g_1}\,z+{\kappa^6\over 16c^2}\,{g_2\over g_1}\,z^2,\eqno(4.16a)$$
$$g_2\sim 1+4\,{\kappa^{3/2}\over c^{1/2}}\sqrt{g_2\over g_1}\,z+{g_2\over g_1}\,z^2.\eqno(4.16b)$$
If we put $r=\sqrt{g_1/g_2}$ and introduce a small parameter $\epsilon=(\kappa^3/c)^{1/2}$, then we deduce from (4.16) that
$$r^2\sim {r^2+\epsilon^3zr+{\epsilon^4\over 16}\,z^2\over r^2+4\epsilon zr+z^2}. \eqno(4.17)$$
It turns out that $r$ satisfies the quartic equation
$$r^4+4\epsilon zr^3+(z^2-1)r^2-\epsilon^3zr-{\epsilon^4\over 16}\,z^2+O(\epsilon^5)=0. \eqno(4.18)$$
\par
The expression of $u^2$ from (3.3a) can be expressed in terms of $r$ as $u^2=\kappa^3+({\rm ln}\, r)_\tau$. We rewrite the $\tau$-derivative
 in accordance with the relation $\partial/\partial\tau=\partial/\partial t+u^2\,\partial/\partial x$ which follows from (2.1b)
and then solve the resultant equation with respect to $u^2$, giving rise to an important relation
$$u^2={r_t+\kappa^3r\over r-r_x}. \eqno(4.19)$$
Differentiating (4.18) by $t$ and $x$, respectively and using the relations $z_t=-cz, z_x=z$ which stems simply from the second expression of (4.15) for $z$, we obtain
$$r_t\sim {2cz^2r^2+4\epsilon czr^3-\epsilon^3czr-{\epsilon^4\over 8}\,cz^2\over 4r^3+2(z^2-1)r+12\epsilon zr^2-\epsilon^3z},\eqno(4.20a)$$
$$r_x\sim {-2z^2r^2-4\epsilon zr^3+\epsilon^3zr+{\epsilon^4\over 8}\,cz^2\over 4r^3+2(z^2-1)r+12\epsilon zr^2-\epsilon^3z}.\eqno(4.20b)$$
Next, if we introduce (4.20) into (4.19) and use (4.18) to eliminate a term $r^4$, we arrive at an approximate expression of $u^2$ in terms of $r$:
$$u^2\sim {cz(2zr^2+4\epsilon r^3-\epsilon^3r-{\epsilon^4\over 8}\,z)
\over 2r^2+2\epsilon^3zr+{1\over 8}\,\epsilon^4z^2}. \eqno(4.21)$$
\par
The last step is to solve  equation (4.18) for $r$ and then substitute the solution into (4.21).  The key feature of our method is that
we need not solve the equation exactly and instead require only the approximate solution. To perform this procedure, we expand $r$ in powers of $\epsilon$ as
$r=\sum_{n=0}^\infty r_n\epsilon^n\ (r_n>0)$ and subsitute this into (4.18). Comparing the coefficients of $\epsilon^n\ (n=0, 1, ...)$, we obtain a system of algebraic
equations for $r_n$. Thus, for $r_0\not=0$, the first three equations arising from the system are found to be as
$$r_0^4+(z^2-1)r_0^2=0, \eqno(4.22a)$$
$$\{4r_0^3+2(z^2-1)r_0\}r_1+4r_0^3z=0, \eqno(4.22b)$$
$$\{4r_0^3+2(z^2-1)r_0\}r_2+(6r_0^2+z^2-1)r_1^2+12zr_0^2r_1=0. \eqno(4.22c)$$
For $r_0=0$, on the other hand,  equations (4.22a) and (4.22b) are satisfied automatically and the first three of  nontrivial equations read
$$(z^2-1)r_1^2=0, \eqno(4.23a)$$
$$2(z^2-1)r_1r_2=0, \eqno(4.23b)$$
$$2(z^2-1)r_1r_3+(z^2-1)r_2^2+r_1^4+4zr_1^3-zr_1-{z^2\over 16}=0. \eqno(4.23c)$$
\begin{figure}[t]
\begin{center}
\includegraphics[width=10cm]{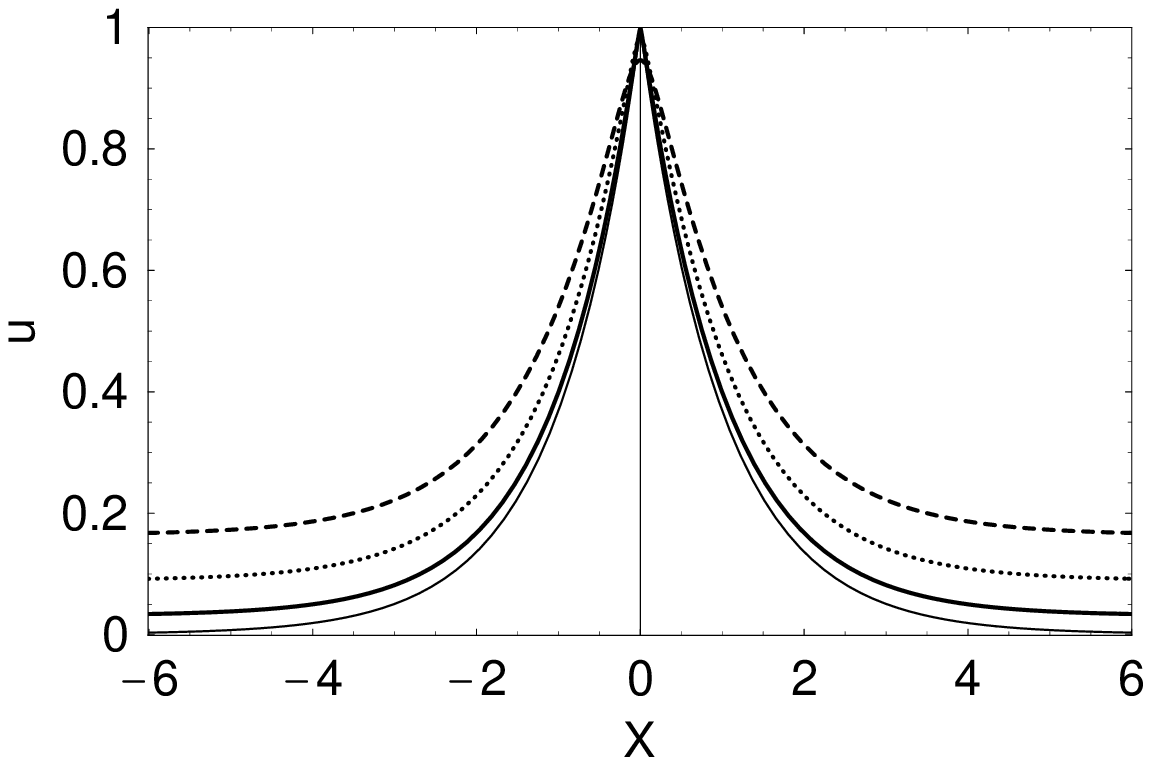}
\end{center}
\noindent {{\bf Figure 3.}\ The peakon limit of  the smooth soliton with $c=1$. $\kappa=0.3$\,(dashed curve), $\kappa=0.2$\,(dotted curve), 
$\kappa=0.1$\,(bold solid curve), $\kappa=0.01$\,(thin solid curve).}
\end{figure}
Thus, if $0<z\leq 1\ (x-ct-x_0\leq 0)$, then $r_0=\sqrt{1-z^2}$ from (4.22a) which, substituted into (4.21), yields
  the limiting form of $u$ when $\epsilon\rightarrow 0$
$$u\sim \sqrt{c}z=\sqrt{c}\,{\rm e}^{x-ct-x_0}. \eqno(4.24a)$$
If, on the other hand,  $z>1\ (x-ct-x_0>0)$, then from (4.23), $r_0=r_1=0$ and $r_2=(1/4)z/\sqrt{z^2-1}$. 
Since then $r\sim O(\epsilon^2)$, we can see that both the numerator and denominator of (4.21) have a leading term of order $\epsilon^4$, which gives a limiting form of $u$ as $\epsilon\rightarrow 0$
$$u\sim \sqrt{c}\,z\sqrt{{r_2^2-{1\over 16}\over r_2^2+{z^2\over 16}}}=\sqrt{c}z^{-1}=\sqrt{c}\,{\rm e}^{-(x-ct-x_0)}. \eqno(4.24b)$$
If we combine (4.24a) with (4.24b), we see that the resulting expression of $u$ coincides perfectly with  the  peakon solution (4.13). \par
The passage to the peakon solution described above is illustrated in figure 3 for four distinct values  of  $\kappa$. 
 We can observe that  the profile drawn by the thin solid  curve fits very well with the peakon solution (4.13) with $c=1$. 
This provides a numerical evidence for the validity of the limiting procedure developed here.
\par
\bigskip
\begin{figure}[t]
\begin{center}
\includegraphics[width=10cm]{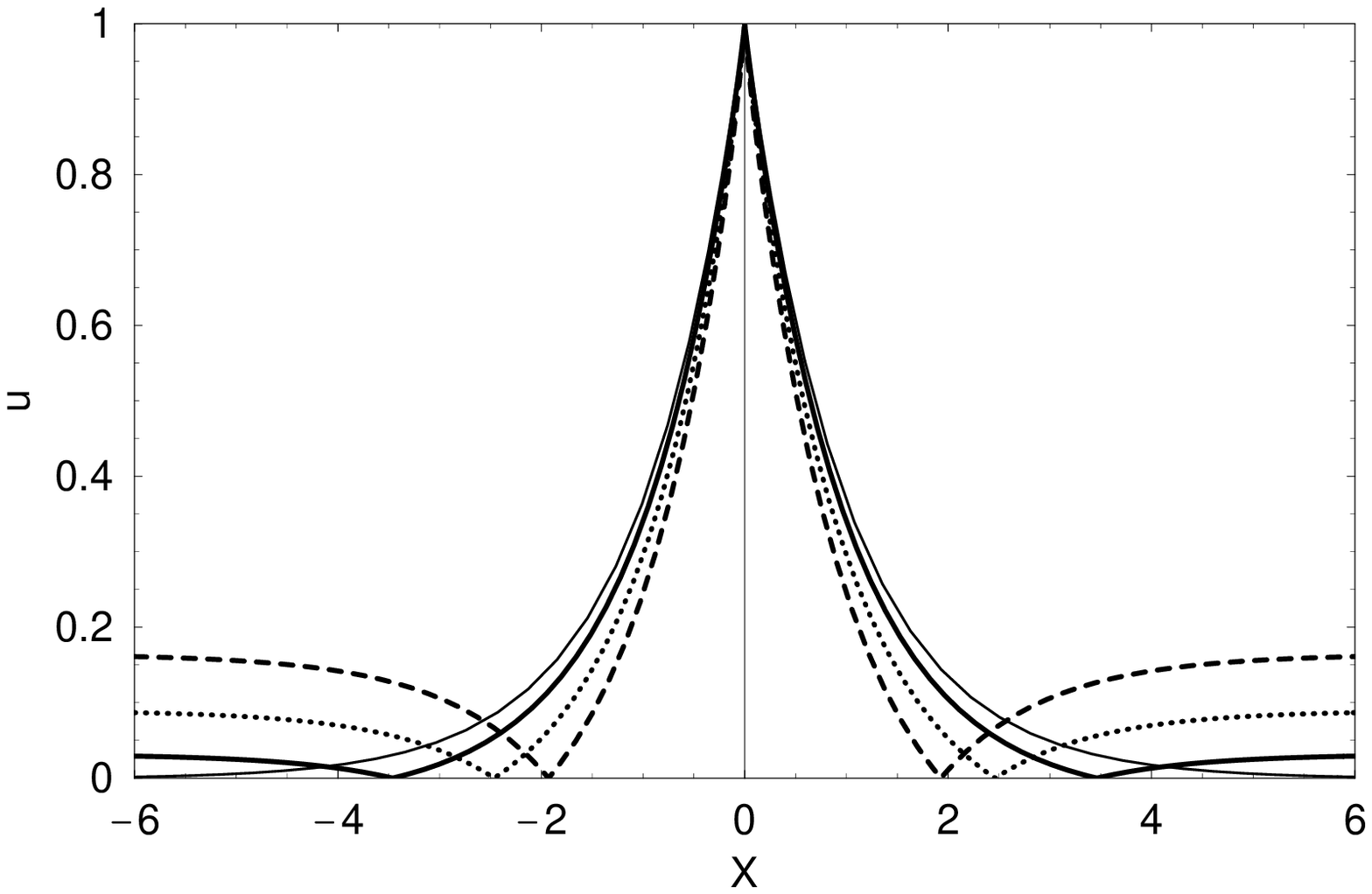}
\end{center}
\noindent {{\bf Figure 4.}\ The peakon limit of  the singular soliton with $c=1$. $\kappa=0.3$\,(dashed curve), $\kappa=0.2$\,(dotted curve), 
$\kappa=0.1$\,(bold solid curve), $\kappa=0.01$\,(thin solid curve).}
\end{figure}
\noindent {\bf Remark 4.1.}\ The asymptotic method employed here for the  smooth one-soliton solution can be applied as well to the general $N$-soliton solution.
The resulting limiting form of the solution should coincide with the $N$-peakon solution which has been obtained by means of the IST [3]. The detailed analysis
will be reported elsewhere. \par
\bigskip
\noindent {\bf Remark 4.2.}\ The peakon limit of the singular soliton can be performed in the same way whereby we use the tau-functions $g_1$ and $g_2$ from (4.14)
with $\xi$ replaced by $\xi+\pi i$. Note that this recipe is equivalent to change the sign of the parameter $\epsilon$ in (4.17).
The subsequent calculation parallels the smooth soliton case. As a result, the limiting form of the singular soliton as $\epsilon\rightarrow 0$ (or equivalently $\kappa\rightarrow 0$) 
is precisely  given by (4.24),
reproducing the peakon solution (4.13).
Specifically, the cusp at the origin turns out to be a peak whereas the two peak positions   $X=\pm X_0$  move to infinity as evidenced from (4.10) by taking the limit $\alpha\rightarrow 1$.
The passage to the peakon limit  is illustrated in figure 4 for four distinct values of $\kappa$. 
 \par
\leftline{\it 4.2. Two-soliton solution}\par
\noindent The tau-functions $g_1$ and $g_2$ for the two-soliton solution are given by (3.1) and (3.2) (or from (2.20), (2.22) and (3.5) ) with $N=2$.  They read
$$g_1=1+2b_1{\rm e}^{\xi_1}+2b_2{\rm e}^{\xi_2}+(a_1b_1)^2{\rm e}^{2\xi_1}+(a_2b_2)^2{\rm e}^{2\xi_2}+2\nu b_1b_2{\rm e}^{\xi_1+\xi_2}
+2\delta b_2(a_1b_1)^2b_2{\rm e}^{2\xi_1+\xi_2}$$
$$+2\delta b_1(a_2b_2)^2{\rm e}^{\xi_1+2\xi_2}+\delta^2(a_1a_2b_1b_2)^2{\rm e}^{2\xi_1+2\xi_2}, \eqno(4.25a)$$
$$g_2=1+{2\over a_1^2 b_1}{\rm e}^{\xi_1}+{2\over a_2^2b_2}{\rm e}^{\xi_2}+{1\over (a_1b_1)^2}{\rm e}^{2\xi_1}+{1\over (a_2b_2)^2}{\rm e}^{2\xi_2}+{2\nu\over (a_1a_2)^2b_1b_2}{\rm e}^{\xi_1+\xi_2}
+{2\delta\over (a_1a_2)^2b_1^2b_2}{\rm e}^{2\xi_1+\xi_2}$$
$$+{2\delta\over (a_1a_2)^2b_1b_2^2}{\rm e}^{\xi_1+2\xi_2}+{\delta^2\over (a_1a_2b_1b_2)^2}{\rm e}^{2\xi_1+2\xi_2}, \eqno(4.25b)$$
where 
$$\xi_i=k_i(y-\tilde c_i\tau-y_{i0}),\quad \tilde c_i={3\kappa^4\over 1-(\kappa k_i)^2},\quad (i=1, 2),\eqno(4.25c)$$
$$a_i=\sqrt{1-{(\kappa k_i)^2\over 4}\over 1-(\kappa k_i)^2},\qquad b_i={1-\kappa k_i\over 1+{\kappa k_i\over 2}}, \quad (i=1, 2), \eqno(4.25d)$$
$$\delta={(k_1-k_2)^2[(k_1^2-k_1k_2+k_2^2)\kappa^2-3]\over (k_1+k_2)^2[(k_1^2+k_1k_2+k_2^2)\kappa^2-3]},\qquad
\nu={(2k_1^4-k_1^2k_2^2+2k_2^4)\kappa^2-6(k_1^2+k_2^2)\over (k_1+k_2)^2[(k_1^2+k_1k_2+k_2^2)\kappa^2-3]}. \eqno(4.25e)$$
 Note that the relation ${\rm e}^{-\phi_i}=a_ib_i$ holds by (2.21) and (4.25d). \par
 Here, we restrict our consideration to the interaction process of two smooth solitons. The other types of the interactions such as soliton-peakon interaction and smooth soliton-singular soliton interaction will not
 be considered. \par
 Let $c_i\ (i=1, 2)$ be the velocity of the $i$the soliton in the $(x, t)$ coordinate system
and assume that $0<c_2<c_1$ which, by (4.2c), is equivalent to imposing  the conditions $0<\kappa k_2<\kappa k_1<1$.
\par
First, we take the limit  $t\rightarrow -\infty$ with $\xi_1$ being fixed. In this limit, $\xi_2\rightarrow-\infty$. 
Then, the leading-order asymptotics of the tau-functions are given by
$$g_1\sim 1+2b_1{\rm e}^{\xi_1}+(a_1b_1)^2{\rm e}^{2\xi_1}, \eqno(4.26a)$$
$$g_2\sim 1+{2\over a_1^2b_1}{\rm e}^{\xi_1}+{1\over (a_1b_1)^2}{\rm e}^{2\xi_1}. \eqno(4.26b)$$
The asymptotic form of the solution follows from (3.3) and (4.26). It can be written as
$$u\sim u_1(\xi_1), \eqno(4.27a)$$
$$ x-c_1t-x_{10}\sim {\xi_1\over\alpha_1}+{1\over 2}\,{\rm ln}\left({\tanh^2{\xi_1\over 2}-{2\over\alpha_1}\tanh\,{\xi_1\over2}+{4-\alpha_1^2\over 3\alpha_1^2}
\over \tanh^2{\xi_1\over 2}+{2\over\alpha_1}\tanh\,{\xi_1\over2}+{4-\alpha_1^2\over 3\alpha_1^2}}\right), \eqno(4.27b)$$
where where $u_1=u(\xi)$ is the one-soliton solution given by (4.7a) and
$$c_1={\tilde c_1\over \kappa}+\kappa^3={\kappa^3(4-\alpha_1^2)\over 1-\alpha_1^2},\qquad \alpha_1=\kappa k_1. \eqno(4.27c)$$
\par
In the limit $t\rightarrow +\infty$,  on the other hand, $\xi_2\rightarrow +\infty$. The expressions corresponding to (4.26) and (4.27) read
$$g_1\sim (a_2b_2)^2{\rm e}^{2\xi_2}\left(1+2\delta b_1{\rm e}^{\xi_1}+\delta^2(a_1b_1)^2{\rm e}^{2\xi_1}\right), \eqno(4.28a)$$
$$g_2\sim {1\over (a_2b_2)^2}{\rm e}^{2\xi_2}\left(1+{2\delta \over a_1^2b_1}{\rm e}^{\xi_1}+{\delta^2\over (a_1b_1)^2}{\rm e}^{2\xi_1}\right). \eqno(4.28b)$$
$$u\sim u_1\left(\xi_1+\delta_1^{(+)}\right), \eqno(4.29a)$$
$$ x-c_1t-x_{10}\sim {\xi_1\over\alpha_1}+{1\over 2}\,{\rm ln}\left({\tanh^2{1\over 2}(\xi_1+\delta_1^{(+)})-{2\over\alpha_1}\tanh\,{1\over 2}(\xi_1+\delta_1^{(+)})+{4-\alpha_1^2\over 3\alpha_1^2}
\over \tanh^2{1\over 2}(\xi_1+\delta_1^{(+)})+{2\over\alpha_1}\tanh\,{1\over 2}(\xi_1+\delta_1^{(+)})+{4-\alpha_1^2\over 3\alpha_1^2}}\right)$$
$$ +2\,{\rm ln}(a_2b_2), \eqno(4.29b)$$
where $\delta_1^{(+)}={\rm ln}\,\delta$.\par
Let $x_{ic}$ be the center position of the $i$th soliton. Then, as $t\rightarrow -\infty$, we find
$$x_{1c}\sim c_1t+x_{10},\quad (\xi_1=0). \eqno(4.30)$$
As $t\rightarrow +\infty$, on the other hand, $x_{1c}$ reads
$$x_{1c}\sim c_1t+x_{10}-{{\rm ln}\,\delta\over \alpha_1}+2\,{\rm ln}(a_2b_2),\quad (\xi_1=-\delta_1^{(+)}). \eqno(4.31)$$
\par
The above analysis shows that the asymptotic state of the solution for large time is represented by a superposition of two single solitons in the
rest frame of reference. The net effect of the interaction between solitons is the phase shift, which we shall now evaluate. To this end,
we define the phase shift of the $i$th soliton by 
$$\Delta_i=x_{ic}(t\rightarrow+\infty)-x_{ic}(t\rightarrow-\infty),\quad (i=1, 2). \eqno(4.32)$$
Then, we see from (4.25), (4.30) and (4.31) that the large soliton suffers a phase shift
$$\Delta_1=-{1\over \kappa k_1}\,{\rm ln}\left[{(k_1-k_2)^2\{(k_1^2-k_1k_2+k_2^2)\kappa^2-3\}\over (k_1+k_2)^2\{[(k_1^2+k_1k_2+k_2^2)\kappa^2-3\}}\right]
-{\rm ln}\left[{\left(1+{\kappa k_2\over 2}\right)\left(1+\kappa k_2\right)\over \left(1-{\kappa k_2\over 2}\right)\left(1-\kappa k_2\right)}\right]. \eqno(4.33)$$
By the similar asymptotic analysis, the phase shift of the small soliton is found to be as
$$\Delta_2={1\over \kappa k_2}\,{\rm ln}\left[{(k_1-k_2)^2\{(k_1^2-k_1k_2+k_2^2)\kappa^2-3\}\over (k_1+k_2)^2\{[(k_1^2+k_1k_2+k_2^2)\kappa^2-3\}}\right]
+{\rm ln}\left[{\left(1+{\kappa k_1\over 2}\right)\left(1+\kappa k_1\right)\over \left(1-{\kappa k_1\over 2}\right)\left(1-\kappa k1\right)}\right]. \eqno(4.34)$$
\par
It is interesting that the above formulas coincide formally with those of the two-soliton solution of the DP equation. 
In the latter case, the parameter $\kappa^3$ is the coefficient of the linear dispersive term $u_x$.
See formula (4.37) of [6].  We can see that there exists a critical curve along which $\Delta_1=\Delta_2$ and beyond which $\Delta_1<\Delta_2$, implying that the
phase shift of the small soliton is greater than that of the large soliton. Such a phenomenon has never been observed in the interaction process  of solitons
for the Korteweg-de Vries and SWW equations.
 \par
\bigskip
\begin{figure}[t]
\begin{center}
\includegraphics[width=16cm]{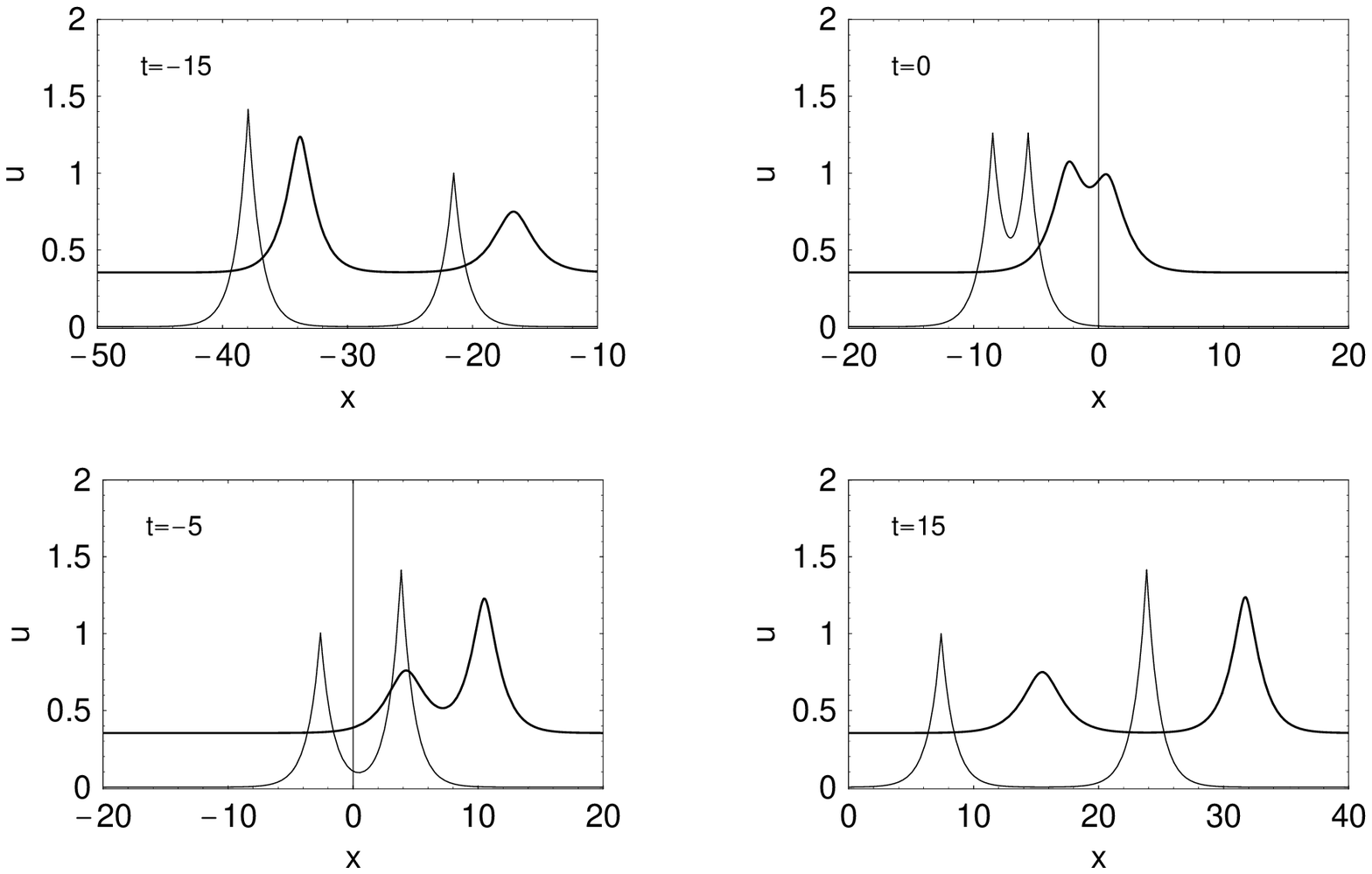}
\end{center}
\noindent {{\bf Figure 5.}\ The profile of the smooth two-soliton solution ($\kappa=0.5$, bold solid curve) and its peakon limit ($\kappa=0.01$, thin solid curve) with $c_1=2, c_2=1$ and $y_{10}=y_{20}=0$.}
\end{figure}
Last, we address the peakon limit of the formula for the phase shift. To this end, we take the limit $\kappa\rightarrow 0$ with $c_1$ and $c_2$ being fixed. 
Substituting the expansion of $\alpha_i(=\kappa k_i)$ for small $\kappa$, $\alpha_i=1-(3/2c_i)\kappa^3+O(\kappa^6),\ (i=1, 2)$ into (4.33) and (4.34), we obtain the following limiting form
of the phase shift:
$$\Delta_1={\rm ln}\left[{c_1(c_1+c_2)\over (c_1-c_2)^2}\right], \qquad \Delta_2={\rm ln}\left[{(c_1-c_2)^2\over c_2(c_1+c_2)}\right]. \eqno(4.35)$$
This result reproduces the formulas for the phase shift of the two-pekaon solution of the Novikov equation [3]. We recall that they coincide formally with the
the corresponding formulas for the two-peakon solution of the DP equation [6]. \par
The profile of the smooth two-soliton solution and its limiting form for small $\kappa$ are depicted in figure 5 for four distinct values of $t$.
 This figure obviously shows the solitonic behaviour of the solution. 
We can observe that the amplitudes of the large and small peakons are $1.41$ and $1.0$, respectively which are in accordance with the those evaluated from the one-peakon solution (4.13).
\par
\bigskip
\leftline{\it 4.3.\ N-soliton solution}\par
\noindent Here, we discribe  the asymptotic behaviour of the general $N$-soliton solution. 
Since the analysis almost parallels  that of the two-soliton case, we outline the result.
To this end, we order the magnitude of the velocity of each soliton as
$0<c_N<c_{N-1}<...<c_1$ by imposing the conditions  $0<\kappa k_N<\kappa k_{N-1}<...<\kappa k_1<1$. \par
We first take the limit $t \rightarrow-\infty$ with the coordinate $\xi_i$ of the $i$th soliton being fixed. Then, $\xi_1, \xi_2, ..., \xi_{i-1}\rightarrow+\infty$,
and $\xi_{i+1}, \xi_{i+2}, ..., \xi_N\rightarrow-\infty$.   We employ (2.20), (2.22) and (3.5) to derive the followng  asymptotic forms of $g_1$ and $g_2$  
$$g_1\sim \beta_i^2\,{\rm exp}\left[2\sum_{j=1}^{i-1}(\xi_j-\phi_j)\right]\left[1+{4(1-\alpha_i)\over 2+\alpha_i}\,{\rm e}^{\xi_i+\delta_i^{(-)}}
+{\left(1-{\alpha_i\over 2}\right)\left(1-\alpha_i\right)\over \left(1+{\alpha_i\over 2}\right)\left(1+\alpha_i\right)}\,{\rm e}^{2(\xi_i+\delta_i^{(-)})}\right], \eqno(4.36a)$$
$$g_2\sim \beta_i^2\,{\rm exp}\left[2\sum_{j=1}^{i-1}(\xi_j+\phi_j)\right]\left[1+{4(1+\alpha_i)\over 2-\alpha_i}\,{\rm e}^{\xi_i+\delta_i^{(-)}}
+{\left(1+{\alpha_i\over 2}\right)\left(1+\alpha_i\right)\over \left(1-{\alpha_i\over 2}\right)\left(1-\alpha_i\right)}\,{\rm e}^{2(\xi_i+\delta_i^{(-)})}\right], \eqno(4.36b)$$
where
$$\delta_i^{(-)}=\sum_{j=1}^{i-1}\,\gamma_{ij}, 
\qquad \beta_i={\rm exp}\left[\sum_{1\leq j<k\leq i-1}\gamma_{jk}\right], \qquad \alpha_i=\kappa k_i. \qquad (i=1, 2, ..., N). \eqno(4.36c)$$
The asymptotic form of the solution is computed by using (3.3) and (4.36) to give
$$u\sim u_1(\xi_i+\delta_i^{(-)}), \eqno(4.37a)$$
$$x-c_it-x_{i0}\sim {\xi_i\over\alpha_i}+{1\over 2}\,{\rm ln}\left({\tanh^2{1\over 2}(\xi_1+\delta_i^{(-)})-{2\over\alpha_i}\tanh\,{1\over 2}(\xi_i+\delta_i^{(-)})+{4-\alpha_i^2\over 3\alpha_i^2}
                    \over \tanh^2{1\over 2}(\xi_i+\delta_1^{(-)})+{2\over\alpha_i}\tanh\,{1\over 2}(\xi_i+\delta_1^{(-)})+{4-\alpha_i^2\over 3\alpha_i^2}}\right)
                     -2\sum_{j=1}^i\phi_j. \eqno(4.37b)$$
 \par
In the limit $t\rightarrow+\infty$, on the other hand, 
$\xi_1, \xi_2, ..., \xi_{i-1}\rightarrow-\infty$,
and $\xi_{i+1}, \xi_{i+2}, ..., \xi_N\rightarrow+\infty$ and the expressions corresponding to (4.36) and (4.37) take the form
$$g_1\sim {\beta_i^\prime}^2\,{\rm exp}\left[2\sum_{j=i+1}^{N}(\xi_j-\phi_j)\right]\left[1+{4(1-\alpha_i)\over 2+\alpha_i}\,{\rm e}^{\xi_i+\delta_i^{(+)}}
+{\left(1-{\alpha_i\over 2}\right)\left(1-\alpha_i\right)\over \left(1+{\alpha_i\over 2}\right)\left(1+\alpha_i\right)}\,{\rm e}^{2(\xi_i+\delta_i^{(+)})}\right], \eqno(4.38a)$$
$$g_2\sim {\beta_i^\prime}^2\,{\rm exp}\left[2\sum_{j=i+1}^{N}(\xi_j+\phi_j)\right]\left[1+{4(1+\alpha_i)\over 2-\alpha_i}\,{\rm e}^{\xi_i+\delta_i^{(+)}}
+{\left(1+{\alpha_i\over 2}\right)\left(1+\alpha_i\right)\over \left(1-{\alpha_i\over 2}\right)\left(1-\alpha_i\right)}\,{\rm e}^{2(\xi_i+\delta_i^{(+)})}\right], \eqno(4.38b)$$
where
$$\delta_i^{(+)}=\sum_{j=i+1}^{N}\,\gamma_{ij}, \qquad \beta_i^\prime={\rm exp}\left[\sum_{i+1\leq j<k\leq N}\gamma_{jk}\right], \eqno(4.38c)$$
and
$$u\sim u_1(\xi_i+\delta_i^{(+)}), \eqno(4.39a)$$
$$x-c_it-x_{i0}\sim {\xi_i\over\alpha_i}+{1\over 2}\,{\rm ln}\left({\tanh^2{1\over 2}(\xi_1+\delta_i^{(+)})-{2\over\alpha_i}\tanh\,{1\over 2}(\xi_i+\delta_i^{(+)})+{4-\alpha_i^2\over 3\alpha_i^2}
                    \over \tanh^2{1\over 2}(\xi_i+\delta_1^{(+)})+{2\over\alpha_i}\tanh\,{1\over 2}(\xi_i+\delta_1^{(+)})+{4-\alpha_i^2\over 3\alpha_i^2}}\right)
                     -2\sum_{j=i+1}^N\phi_j. \eqno(4.39b)$$
\par 
The phase shift of the $i$th soliton is evaluated from (4.37) and (4.39). It reads
$$\Delta_i={1\over \kappa k_i}\sum_{j=1}^{i-1}{\rm ln}\left[{(k_i-k_j)^2\{(k_i^2-k_ik_j+k_j^2)\kappa^2-3\}\over (k_i+k_j)^2\{[(k_i^2+k_ik_j+k_j^2)\kappa^2-3\}}\right]$$
$$-{1\over \kappa k_i}\sum_{j=i+1}^{N}{\rm ln}\left[{(k_i-k_j)^2\{(k_i^2-k_ik_j+k_j^2)\kappa^2-3\}\over (k_i+k_j)^2\{(k_i^2+k_ik_j+k_j^2)\kappa^2-3\}}\right]
+\sum_{j=1}^{i-1}{\rm ln}\left[{\left(1+{\kappa k_i\over 2}\right)\left(1+\kappa k_i\right)\over \left(1-{\kappa k_i\over 2}\right)\left(1-\kappa k_i\right)}\right]$$
$$-\sum_{j=i+1}^N{\rm ln}\left[{\left(1+{\kappa k_i\over 2}\right)\left(1+\kappa k_i\right)\over \left(1-{\kappa k_i\over 2}\right)\left(1-\kappa k_i\right)}\right], \qquad (i=1, 2, ..., N). \eqno(4.40)$$
\par
The above formulas for the phase shift clearly show that each soliton has a pairwise interaction with other solitons.
For the special case of $N=2$, they reduce to the corresponding formulas for the two-soliton solution (4.33) and (4.34).
The first two terms on the right-hand side of (4.40) coincide
with the phase shift of the $i$th soliton of the SWW equation whereas the last two additional terms stem from the coordinate transformation (2.1). A novel feature of the phase shift would
appear due to the latter terms. Note, remarkably that the formulas (4.40) are formally the same as those of the DP equation. See formulas (4.52) of [6] as well as formulas (4.11) of [7].
\par
The peakon limit of (4.40) can be carried out straightforwardly to give the formulas
$$\Delta_i=\sum_{j=1}^{i-1}{\rm ln}\left[{(c_i-c_j)^2\over c_i(c_i+c_j)}\right]-\sum_{j=i+1}^N{\rm ln}\left[{(c_i-c_j)^2\over c_i(c_i+c_j)}\right],\qquad (i=1, 2, ..., N), \eqno(4.41)$$
which reproduce the corresponding formulas for the $N$-peakon solution of the Novikov equation [3] and they coincide formally with those of the DP equation [6]. \par
\bigskip
\leftline{\bf 5. Conservation laws}\par
\bigskip
\noindent The existence of an infinite number of conservation laws is a common feature to integrable nonlinear PDEs.  The conservation laws of the Novikov
equation have been derived by means of the IST [2]. Actually, this has been established by making use of the Lax pair (2.7) rewritten in terms of the variables $x$ and $t$. 
Here, we present an alternative method on the basis of the relation (2.19) which
connects solutions of the Novikov equation and those of the SWW equation (2.16). To this end, we substitute $U$ from (2.9) into (2.19) to obtain
$$q={1\over 3}\left({p_{yy}\over 2p}-{3\over 4}\,{p_y^2\over p^2}-p^2+{1\over\kappa^2}\right). \eqno(5.1a)$$
Upon substituting the defintion of $p$ i.e., $p=m^{-2/3}$ and rewriting the result  by the variables $x$ and $t$  in accordance with (2.1b),  relation (5.1a)
can be put into the form
$$q=-{1\over 3}\left\{m^{-4/3}\left(1-{4\over 9}m^{-2}m_x^2+{1\over 3}m^{-1}m_{xx}\right)-{1\over\kappa^2}\right\}. \eqno(5.1b)$$
\par
The above relation enables us to construct conservation lows of the Novikov equation from those of the SWW equation, as we shall now demonstrate.  The latter equation  has local
conservation laws of the form
$$w_{n,\tau}=j_{n,y},\qquad (n=0, 1, 2, ...), \eqno(5.2)$$
where the conserved density $w_n$ and associated flux $j_n$ are polynomials of $q$ and its $y$-derivatives. We rewrite (5.2) in terms of the variables $x$ and $t$ 
whereby we employ equation (1.1) modified in the form $(m^{2/3})_t+(m^{2/3}u^2)_x=0$. The resulting expression reads
$$(m^{2/3}w_n)_t=(j_n-m^{2/3}u^2w_n)_x \eqno(5.3)$$
It turns out that the quantities
$$I_n=\int_{-\infty}^\infty (m^{2/3}w_n-\kappa w_{n0})dx,\quad 
(n=0, 1, 2, ...), \eqno(5.4)$$
are coonserved, where $w_{n0}$ is the boundary value of $w_n$ as $|x|\rightarrow \infty$. 
The  conserved densities $w_n$ for the SWW equation has been obtained by means of the B\"acklund transformation.  According to  [10], for instance, the first three of them can be written as
$$w_0=1,\quad w_1=q, \quad w_2=q^3-q_y^2-{1\over\kappa^2}q^2,\quad w_3=q^4-3qq_y^2+{1\over 3}q_{yy}^2-{2\over 3\kappa^2}\left(q^3-{1\over 2}q_y^2\right). \eqno(5.5)$$
Note that the terms expressed by a perfect derivative $(W_n)_y$ have been dropped in (5.5) since they become zero after integrating with respect to $x$.
Actually, we can show by using (2.1b) that $\int_{-\infty}^\infty m^{2/3}(W_n)_ydx=\int_{-\infty}^\infty(W_n)_xdx=0.$
 The conservation laws constructed in this way have very lengthy expressions  and hence  we quote the first two of them here:
$$I_0=\int_{-\infty}^\infty(m^{2/3}-\kappa)dx, \qquad I_1=\int_{-\infty}^\infty\left[m^{-2/3}\left(1-{4\over 9}m^{-2}m_x^2+{1\over 3}m^{-1}m_{xx}\right)-{1\over\kappa}\right]dx. \eqno(5.6)$$
Recall that $I_0$ follows directly from (1.1) and $I_1$ coincides with the corresponding law obtained by the IST [2].  \par
\bigskip
\leftline{\bf 6. Concluding remarks}\par
\bigskip
\noindent In this paper, we have developed a systematic method for solving the Novikov equation. In particular, we have obtained the parametric representation of the
$N$-soliton solution in terms of the tau-functions associated with those of the SWW equation. A detailed analysis of the structure of the solutions 
reveals various new features which have never seen in existing soliton solutions. 
Although there are many more things to be explored for the Novikov equation, the present paper will shed some light on the subject.
\par
In conclusion, it will be worthwhile to compare the results
obtained here with those of an another version of the modified CH equation, which is given by
$$m_t+[m(u^2-u_x^2)]_x=0,\qquad m=u-u_{xx}. \eqno(6.1)$$
The above equation has been derived by several researchers as an integrable generalization of the CH equation  and attracted a lot of interest in recent years [11-18].
It exhibits smooth  solitons [17, 18] on a constant background as well as singular peaked solitons with W-shaped profile and cusp solitons [14, 15]. 
Although both equations have cubic nonlinearities, their structure differs substantially as observed by comparing their Lax representations.
Actually, the spatial part of  the Lax pair for the Novikov equation is the third-order ODE [1, 2] 
whereas the corresponding one for the modified CH equation is the second-order ODE [14]. 
Generally speaking, the former problem is more difficult to analyze than the latter one. This would be a reason why 
the IST has not still succeeded in obtaining smooth multisoliton solutions of the Novikov equation. In this respect, however we recall that the construction of
peakon sulutions has gone ahead as in the cases of the CH and DP equations [2, 3]. 
\par
 The parametric representation for the bright $N$-soliton solution of equation (6.1) has been
obtained quite recently by means of a reciprocal transformation [18]. Specifically, it has been demonstrated that the transformed equation is closely related to the SWW equation
introduced by Ablowitz et al [19]
$$q_\tau+2\kappa^3\,q_y-4\kappa^2qq_\tau +2\kappa^2\,q_y\int_y^\infty q_\tau dy-\kappa^2q_{\tau yy}=0, \qquad q=q(y, \tau). \eqno(6.2)$$
It is well-known that equation (6.2) has a different mathematical structure from that of the SWW equation of Hirota and Satsuma [5, 20]. 
The $N$-soliton solution of equation (6.1) is represented in terms of the tau-functions associated with the $N$-soliton solution of the  equation (6.2)  [18].
This intriguing feature is similar to that between the CH and DP equations.
Actually, the structure of the soliton solutions of the Novikov (modified CH) equation bears a close resemblance to that of the DP (CH) equation.
 This fact can be seen if one compares the formulas
for the phase shift of solitons, for instance. 
See also a few papers related to the similar approach to that developed here [21-25].
\par
As for the singular solutions, equation (6.1) exhibits both W-shaped solitons with three peaks and cusp solitons whereas the Novikov equation admits the  
solitons with a similar W-shaped profile but with single cusp and double peaks, as shown in  figure 2.  
Another distinct feature is that the peakon solutions of the Novikov equation can be reduced from the smooth soliton solutions as 
the background field tends to zero. However, the similar limiting procedure has not been undertaken for the bright soliton solutions of equation (6.1) [26]. 
This is indeed a very interesting issue to be clarified by further study.

\newpage
\leftline{\bf Appendix. Proof of the bilinear identities} \par
\bigskip
\noindent
In this appendix, we perform the proof of the various identities presented in proposition 2.2 as well as (2.13) for the $N$-soliton solution of the SWW equation.
The proof will be done by mathematical induction similar to that used for the proof of the bilinear identities associated with the $N$-soliton solution of the
DP equation. See section 3 of [7]. We first prove (2.13) and then proceed to (2.23)-(2.26). \par
\bigskip
\leftline{\it A.1. Proof of (2.13)}\par
\noindent Substituting  the tau-function $f$ from (2.20) into (2.13) and using the formula
$$D_\tau^mD_y^n\,{\rm exp}\left[\sum_{i=1}^N\mu_i\xi_i\right]\cdot {\rm exp}\left[\sum_{i=1}^N\nu_i\xi_i\right]$$
$$=\left\{-\sum_{i=1}^N(\mu_i-\nu_i)k_i\tilde c_i\right\}^m\left\{\sum_{i=1}^N(\mu_i-\nu_i)k_i\right\}^n{\rm exp}\left[\sum_{i=1}^N(\mu_i+\nu_i)\xi_i\right],\qquad (m, n=0, 1, 2, ...),$$
 the identity to be proved becomes
$$\sum_{\mu,\nu=0,1}\Biggl[-\left\{\sum_{i=1}^N(\mu_i-\nu_i)k_i\tilde c_i\right \}\left\{\sum_{i=1}^N(\mu_i-\nu_i)k_i\right\}^3-3\kappa^2\left\{\sum_{i=1}^N(\mu_i-\nu_i)k_i\right \}^2$$
$$+{1\over \kappa^2}\left\{\sum_{i=1}^N(\mu_i-\nu_i)k_i\tilde c_i\right\}\left\{\sum_{i=1}^N(\mu_i-\nu_i)k_i\right\}\Biggr]\times$$
$$\times {\rm exp}\left[\sum_{i=1}^N(\mu_i+\nu_i)\xi_i+\sum_{1\leq i<j\leq N}(\mu_i\mu_j+\nu_i\nu_j)\gamma_{ij}\right]=0. \eqno(A.1)$$
\par
Let $P_{m,n}$ be the coefficient of the factor ${\rm exp}\left[\sum_{i=1}^n\xi_i+\sum_{i=n+1}^m2\xi_i\right]\ (1\leq n<m\leq N)$ on the left-hand side of (A.1). Correspondingly, the
summation with respect to $\mu_i$ and $\nu_i$ must be performed under the conditions
$$\mu_i+\nu_i=1\qquad (i=1, 2, ..., n), \qquad \mu_i=\nu_i=1\qquad (i=n+1, n+2, ..., m),$$
$$\mu_i=\nu_i=0\qquad (i=m+1, m+2, ..., N). \eqno(A.2)$$
To simplify the notation, we introduce the new summation indices $\sigma_i$ by the relations $\mu_i=(1+\sigma_i)/2,\ \nu_i=(1-\sigma_i)/2$ for $i=1, 2, ..., n$, where $\sigma_i$ takes either the value $+1$ or $-1$. 
It turns out that $\mu_i\mu_j+\nu_i\nu_j=(1+\sigma_i\sigma_j)/2$. \par
Now, under the conditions (A.2), we deduce that
$$\sum_{1\leq i<j\leq N}(\mu_i\mu_j+\nu_i\nu_j)\gamma_{ij}={1\over 2}\sum_{1\leq i<j\leq n}(1+\sigma_i\sigma_j)\gamma_{ij}+\sum_{i=1}^m\sum_{\substack{j=n+1\\ (j\not=i)}}^m\gamma_{ij}. \eqno(A.3)$$
Using (A.3), $P_{m,n}$ can be written in the form
$$P_{m,n}=\sum_{\sigma=\pm 1} \Biggl[-\left(\sum_{i=1}^n\sigma_ik_i\tilde c_i\right)\left(\sum_{i=1}^n\sigma_ik_i\right)^3-3\kappa^2\left(\sum_{i=1}^n\sigma_ik_i\right)^2$$
$$+{1\over \kappa^2}\left(\sum_{i=1}^n\sigma_ik_i\tilde c_i\right)\left(\sum_{i=1}^n\sigma_ik_i\right)\Biggr]
{\rm exp}\left[{1\over 2}\sum_{1\leq i<j\leq n}(1+\sigma_i\sigma_j)\gamma_{ij}+\sum_{i=1}^m\sum_{\substack{j=n+1\\ (j\not=i)}}^m\gamma_{ij}\right]. \eqno(A.4)$$
The following relation follows from (2.20c) and the definition of $\sigma_i$:
$${\rm exp}\left[{1\over 2}\sum_{1\leq i<j\leq n}(1+\sigma_i\sigma_j)\gamma_{ij}\right]
={(\sigma_ik_i-\sigma_jk_j)^2[(k_i^2-\sigma_i\sigma_jk_ik_j+k_j^2)\kappa^2-3]\over (k_i+k_j)^2[(k_i^2+k_ik_j+k_j^2)\kappa^2-3]}.\eqno(A.5)$$
Introducing (A.5) into (A.4), the identity to be proved reduces, after multiplying $P_{m,n}$ by a factor $\prod_{i=1}^n(1-\kappa^2k_i^2)$ and using the definition  of $\tilde c_i$ from (4.25c),  to
$$P_n(k_1, k_2, ..., k_n)\equiv \sum_{\sigma=\pm 1} \Biggl[-3\kappa^4\left\{\sum_{i=1}^n\sigma_ik_i\prod_{j=1(j\not=i)}^n(1-\kappa^2k_i^2)\right\}
\left(\sum_{i=1}^n\sigma_ik_i\right)^3$$
$$-3\kappa^2\left(\sum_{i=1}^n\sigma_ik_i\right)^2\prod_{j=1}^n(1-\kappa^2k_i^2)
+{3 \kappa^2}\Biggl\{\sum_{i=1}^n\sigma_ik_i\prod_{j=1(j\not=i)}^n(1-\kappa^2k_j^2)\Biggr\}\left(\sum_{i=1}^n\sigma_ik_i\right)\Biggr]\times$$
$$\times\prod_{1\leq i<j\leq n}(\sigma_ik_i-\sigma_jk_j)^2[(k_i^2-\sigma_i\sigma_jk_ik_j+k_j^2)\kappa^2-3]=0, \qquad (n=1, 2, ..., N),\eqno(A.6)$$
where a multiplicative factor independent of the indices $\sigma_i$ has been dropped.
\par
Before proving (A.6), we  first establish the following identity: 
$$\bar P_n(k_1, k_2, ..., k_n)\equiv\sum_{\sigma=\pm 1}\left(\kappa\sum_{i=1}^n\sigma_ik_i\right)\left(1+\kappa\sum_{i=1}^n\sigma_ik_i\right)\left(1+{1\over 2}\kappa\sum_{i=1}^n\sigma_ik_i\right)\times$$
$$\times \prod_{i=1}^n(1-\kappa \sigma_i k_i)\left(1-{\kappa\over 2}\sigma_i k_i\right)\prod_{1\leq i<j\leq n}(\sigma_ik_i-\sigma_jk_j)^2[(k_i^2-\sigma_i\sigma_jk_ik_j+k_j^2)\kappa^2-3]=0,$$
$$\qquad (n=1, 2, ..., N). \eqno(A.7)$$
The proof proceeds by mathematical induction. A direct calculation shows that $\bar P_1=\bar P_2=0.$  Assume that $\bar P_{n-2}=\bar P_{n-1}=0$. Then,
$$\bar P_n\big|_{k_1=0}=\prod_{i=2}^nk_i^2(\kappa^2k_i^2-3)\,\bar P_{n-1}(k_2, k_3, ..., k_n)=0.\eqno(A.8)$$
$$\bar P_n\big|_{k_1=1/\kappa}=0,\eqno(A.9)$$
$$\bar P_n\big|_{k_1=k_2}=-12k_1^2(1-\kappa^2k_1^2)^2\left(1-{1\over 4}\kappa^2k_1^2\right)\times$$
$$\times\prod_{i=3}^n(k_1^2-k_i^2)\left[(k_1^4+k_1^2k_i^2+k_i^4)\kappa^4-6(k_1^2+k_i^2)\kappa^2+9\right]\bar P_{n-2}(k_3, k_4, ..., k_n)=0. \eqno(A.10)$$
Note that (A.9) is proved without use of the assumption of induction. The polynomial $\bar P_n$ is symmetric and even function of $k_i\, (i=1, 2, ..., n)$.
With this fact and (A.8)-(A.10) in mind, we see that $\bar P_n$ can be factored by a polynomial
$$\prod_{i=1}^nk_i^2\left(k_i^2-{1\over\kappa^2}\right)\prod_{1\leq i<j\leq n}(k_i^2-k_j^2)^2,$$
of $k_i\ (i=1, 2, ..., n)$ of degree $2n^2+2n$. On the other-hand, the degree of $\bar P_n$ from (A.7) is $2n^2+3$ at most, which is impossible except for 
$\bar P_n\equiv 0.$ This completes the proof of (A.7).\par
We now prove (A.6) following the same procedure as that used for (A.7).  It is easy to check that $P_1=P_2=0.$  Assume that $P_{n-2}=P_{n-1}=0$. Then,
$$P_n\big|_{k_1=0}=\prod_{i=2}^nk_i^2(\kappa^2k_i^2-3)\,P_{n-1}(k_2, k_3, ..., k_n)=0,\eqno(A.11)$$
$$P_n\big|_{k_1=1/\kappa}=-12\prod_{i=2}^n\left(-{2\over\kappa^2}\right)(1-\kappa^2k_i^2)^2\,\bar P_{n-1}(k_2, k_3, ..., k_n)=0,\eqno(A.12)$$
$$ P_n\big|_{k_1=k_2}=4k_1^2(1-\kappa^2k_1^2)^2
\prod_{i=3}^n(k_1^2-k_i^2)^2\left[(k_1^4+k_1^2k_i^2+k_i^4)\kappa^4-6(k_1^2+k_i^2)\kappa^2+9\right]\times$$
$$\times P_{n-2}(k_3, k_4, ..., k_n)=0. \eqno(A.13)$$
where we have used (A.7) in (A.12). The relations (A.11)-(A.13) together with the symmetry and evenness of $P_n$ in $k_i (i=1, 2, ..., n)$ imply that $P_n$ has a factor
$$\prod_{i=1}^nk_i^2\left(k_i^2-{1\over\kappa^2}\right)\prod_{1\leq i<j\leq n}(k_i^2-k_j^2)^2,$$
whose degree in $k_i\ (i=1, 2, ..., n)$ is $2n^2+2n$ whereas the degree of $P_n$ from (A.6) is $2n^2+2$ at most. This is impossible 
except for $P_n\equiv 0$, completing the proof of (A.6) and hence (2.13). \hspace{\fill}$\Box$ \par
\bigskip
\leftline{\it A.2. Proof of (2.23)}\par
\bigskip
\noindent We can perform the proof by following the procedure used in the proof of (2.13) and hence we describe the outline. In the present case, the identity to be proved can be written in the form
$$Q_n(k_1, k_2, ..., k_n)\equiv  \sum_{\sigma=\pm 1}\Biggl[\left(\sum_{i=1}^n\sigma_ik_i+{2\over\kappa}\right)\prod_{i=1}^n\left\{{1-{\kappa\sigma_ik_i\over 2}\over 1+{\kappa\sigma_ik_i\over 2}}(1-\kappa\sigma_ik_i)^2\right\}$$
$$-{2\over\kappa^3}\Biggl\{\kappa^2\prod_{i=1}^n(1-\kappa^2k_i^2)+3\kappa^4\Biggl(\sum_{i=1}^n\sigma_ik_i\Biggl)\Biggl(\sum_{i=1}^n\sigma_ik_i\prod_{j=1(j\not=i)}^n(1-\kappa^2k_j^2)\Biggr)\Biggr\}\Biggr]\times$$
$$\times\prod_{1\leq i<j\leq n}(\sigma_ik_i-\sigma_jk_j)^2[(k_i^2-\sigma_i\sigma_jk_ik_j+k_j^2)\kappa^2-3]=0, \qquad (n=1, 2, ..., N).\eqno(A.14)$$
Let $Q_n^\prime$ be the first term of $Q_n$ multiplied by a factor $\prod_{i=1}^n\left(1-{1\over 4}\kappa^2k_i^2\right)$, i.e.
$$Q_n^\prime=\sum_{\sigma=\pm 1}\left(\sum_{i=1}^n\sigma_ik_i+{2\over\kappa}\right)\prod_{i=1}^n\left(1-{\kappa\sigma_ik_i\over 2}\right)^2(1-\kappa\sigma_ik_i)^2\times$$
$$\times\prod_{1\leq i<j\leq n}(\sigma_ik_i-\sigma_jk_j)^2[(k_i^2-\sigma_i\sigma_jk_ik_j+k_j^2)\kappa^2-3].$$
 We can show that $Q_n^\prime\big|_{k_1=2/\kappa}=-Q_n^\prime\big|_{k_1=2/\kappa}$ and hence $Q_n^\prime\big|_{k_1=2/\kappa}=0$. 
The polynomial $Q_n^\prime$ is symmetric and even function of $k_i\, (i=1, 2, ..., n)$.
Thus, the polynomial $Q_n^\prime$ 
has a factor $\prod_{i=1}^n(1-{1\over 4}\kappa^2 k_i^2)$, implying  the first term of $Q_n$ is indeed a polynomial. Taking into account this fact, we now start the proof of (A.14). 
A direct calculation shows that $Q_1=Q_2=0.$  Assume that $Q_{n-2}=Q_{n-1}=0$. Then,
$$Q_n\big|_{k_1=0}=\prod_{i=2}^nk_i^2(\kappa^2k_i^2-3)\,Q_{n-1}(k_2, k_3, ..., k_n)=0,\eqno(A.15)$$
$$Q_n\big|_{k_1=1/\kappa}=0, \eqno(A.16)$$
$$ Q_n\big|_{k_1=k_2}=-12k_1^2(1-\kappa^2k_1^2)^3
\prod_{i=3}^n(k_1^2-k_i^2)^2\left[(k_1^4+k_1^2k_i^2+k_i^4)\kappa^4-6(k_1^2+k_i^2)\kappa^2+9\right]\times$$
$$\times Q_{n-2}(k_3, k_4, ..., k_n)=0. \eqno(A.17)$$
The symmetry and evenness of $Q_n$ in $k_i\, (i=1, 2, ..., n)$ as well as the relations (A.15)-(A.17) show that $Q_n$ has a factor
$$\prod_{i=1}^nk_i^2\left(k_i^2-{1\over\kappa^2}\right)\prod_{1\leq i<j\leq n}(k_i^2-k_j^2)^2,$$
whose degree in $k_i\ (i=1, 2, ..., n)$ is $2n^2+2n$. On the other-hand, the degree of $Q_n$ from (A.14) is $2n^2+1$ at most. 
This is impossible except for $Q_n\equiv 0$, completing the proof of (A.14) and hence (2.23). \hspace{\fill}$\Box$ \par
\bigskip
\leftline{\it A.3. Proof of (2.24)}\par
\bigskip
\noindent The identity to be proved reads
$$R_n(k_1, k_2, ..., k_n)\equiv  \sum_{\sigma=\pm 1}\Biggl[\Biggl\{-3\kappa^4\sum_{i=1}^n\sigma_ik_i
\prod_{j=1(j\not=i)}^n(1-\kappa^2k_j^2)+2\kappa^3\prod_{j=1}^n(1-\kappa^2k_i^2)\Biggr\}\times$$
$$\times\prod_{i=1}^n\left\{ {\Bigl(1-{\kappa\sigma_ik_i\over 2}\Bigr)(1-\kappa\sigma_ik_i)^2\over 1+{\kappa\sigma_ik_i\over 2}}\right\}$$
$$-{2\over\kappa^3}\Biggl\{\kappa^6\prod_{j=1}^n(1-\kappa^2k_i^2)^2+\Biggl(3\kappa^4\sum_{i=1}^n\sigma_ik_i\prod_{j=1(j\not=i)}^n(1-\kappa^2k_j^2)\Biggr)^2\Biggr\}\Biggr]\times$$
$$\times \prod_{1\leq i<j\leq n}(\sigma_ik_i-\sigma_jk_j)^2[(k_i^2-\sigma_i\sigma_jk_ik_j+k_j^2)\kappa^2-3]=0, \qquad (n=1, 2, ..., N).\eqno(A.18)$$
The  $R_n$ is  a polynomial in  $k_i\ (i=1, 2, ..., n)$, as shown by an argument similar to that given for $Q_n^\prime$.
A direct calculation shows that $R_1=R_2=0.$  Assume that $R_{n-2}=R_{n-1}=0$. Then,
$$R_n\big|_{k_1=0}=\prod_{i=2}^nk_i^2(\kappa^2k_i^2-3)\,R_{n-1}(k_2, k_3, ..., k_n)=0,\eqno(A.19)$$
$$R_n\big|_{k_1=1/\kappa}=0, \eqno(A.20)$$
$${\partial R_n\over\partial k_1}\big|_{k_1=1/\kappa}=0, \eqno(A.21)$$
$$ R_n\big|_{k_1=k_2}=-12k_1^2(1-\kappa^2k_1^2)^4
\prod_{i=3}^n(k_1^2-k_i^2)^2\left[(k_1^4+k_1^2k_i^2+k_i^4)\kappa^4-6(k_1^2+k_i^2)\kappa^2+9\right]\times$$
$$\times R_{n-2}(k_3, k_4, ..., k_n)=0. \eqno(A.22)$$
Note that (A.20) and (A.21) follow by direct computation without use of the assumption of induction.
The symmetry and evenness of $R_n$ in $k_i\, (i=1, 2, ..., n)$ as well as the relations (A.19)-(A.22) show that $Q_n$ has a factor
$$\prod_{i=1}^nk_i^2\left(k_i^2-{1\over\kappa^2}\right)^2\prod_{1\leq i<j\leq n}(k_i^2-k_j^2)^2,$$
whose degree in $k_i\ (i=1, 2, ..., n)$ is $2n^2+4n$.
On the other-hand, the degree of $R_n$ from (A.18) is $2n^2+2n$ at most. 
This is impossible except for $R_n\equiv 0$, completing the proof of (A.18) and hence (2.24). \hspace{\fill}$\Box$ \par
\bigskip
\leftline{\it A.4. Proof of (2.25)}\par
\bigskip
\noindent The identity to be proved reads
$$S_n(k_1, k_2, ..., k_n)\equiv  \sum_{\sigma=\pm 1}\Biggl[\Biggl\{\Bigl(\sum_{i=1}^n\sigma_ik_i\Bigr)^3
+{6\over\kappa}\Bigl(\sum_{i=1}^n\sigma_ik_i\Bigr)^2+{11\over\kappa^2}\sum_{i=1}^n\sigma_ik_i+{6\over\kappa^3}\Biggr\}\times$$
$$\times\prod_{i=1}^n\left\{\Bigl(1-{\kappa\sigma_ik_i\over 2}\Bigr)(1-\kappa\sigma_ik_i)  \over \Bigl(1+{\kappa\sigma_ik_i\over 2}\Bigr)(1+\kappa\sigma_ik_i)\right\}-{6\over\kappa^3}\Biggr]
 \prod_{1\leq i<j\leq n}(\sigma_ik_i-\sigma_jk_j)^2[(k_i^2-\sigma_i\sigma_jk_ik_j+k_j^2)\kappa^2-3]=0,$$
 $$ \qquad (n=1, 2, ..., N).\eqno(A.23)$$
The $S_n$ can be shown to be a polynomial in $k_i\ (i=1, 2, ..., n)$.
A direct calculation shows that $S_1=S_2=0.$  Assume that $S_{n-2}=S_{n-1}=0$. Then,
$$ S_n\big|_{k_1=0}=\prod_{i=2}^nk_i^2(\kappa^2k_i^2-3)\,S_{n-1}(k_2, k_3, ..., k_n)=0,\eqno(A.24)$$
$$ S_n\big|_{k_1=k_2}=-12k_1^2(1-\kappa^2k_1^2)
\prod_{i=3}^n(k_1^2-k_i^2)^2\left[(k_1^4+k_1^2k_i^2+k_i^4)\kappa^4-6(k_1^2+k_i^2)\kappa^2+9\right]\times$$
$$\times S_{n-2}(k_3, k_4, ..., k_n)=0. \eqno(A.25)$$
The symmetry and evenness of $S_n$ in $k_i\, (i=1, 2, ..., n)$ as well as the relations (A.24)  and (A.25) show that $S_n$ has a factor
$$\prod_{i=1}^nk_i^2\prod_{1\leq i<j\leq n}(k_i^2-k_j^2)^2,$$
whose degree in $k_i\ (i=1, 2, ..., n)$ is $2n^2$.
On the other-hand, the degree of $S_n$ from (A.23) is $2n^2-2n+3$ at most. 
This is impossible except for $S_n\equiv 0$, completing the proof of (A.23) and hence (2.25). \hspace{\fill}$\Box$ \par
\bigskip
\leftline{\it A.5. Proof of (2.26)}\par
\bigskip
\noindent The identity to be proved reads
$$T_n(k_1, k_2, ..., k_n)\equiv  \sum_{\sigma=\pm 1}\Biggl[\Biggl\{-3\kappa^4\Bigl(\sum_{i=1}^n\sigma_ik_i\prod_{j=1(j\not=i)}^n(1-\kappa^2k_i^2)\Bigr)\Bigl(1+{\kappa\over 2}\sum_{i=1}^n\sigma_ik_i\Bigr)^2$$
$$+{\kappa^3\over 2}\Bigl(\kappa\sum_{i=1}^n\sigma_ik_i-1\Bigr)\Bigl(\kappa\sum_{i=1}^n\sigma_ik_i+2\Bigr)\prod_{j=1}^n(1-\kappa^2k_j^2)\Biggr\}
\prod_{i=1}^n\left\{ {\Bigl(1-{\kappa\sigma_ik_i\over 2}\Bigr)(1-\kappa\sigma_ik_i)^2\over 1+{\kappa\sigma_ik_i\over 2}}\right\}$$
$$-{9\kappa^3\over 2}\biggl(\kappa\sum_{i=1}^n\sigma_ik_i\prod_{j=1(j\not=i)}^n(1-\kappa^2k_j^2)\biggr)^2\Bigl(\kappa \sum_{i=1}^n\sigma_ik_i\Bigr)^2
+\kappa^3\biggl(-{1\over 2}\Bigl(\kappa\sum_{i=1}^n\sigma_ik_i\Bigr)^2+1\biggr)\prod_{j=1}^n(1-\kappa^2k_j^2)^2\Biggr]\times$$
$$\times \prod_{1\leq i<j\leq n}(\sigma_ik_i-\sigma_jk_j)^2[(k_i^2-\sigma_i\sigma_jk_ik_j+k_j^2)\kappa^2-3]=0, \qquad (n=1, 2, ..., N).\eqno(A.26)$$
The $T_n$ is a polynomial in $k_i\ (i=1, 2, ..., n)$.
A direct calculation shows that $T_1=T_2=0.$  Assume that $T_{n-2}=T_{n-1}=0$. Then,
$$T_n\big|_{k_1=0}=\prod_{i=2}^nk_i^2(\kappa^2k_i^2-3)\,T_{n-1}(k_2, k_3, ..., k_n)=0,\eqno(A.27)$$
$$T_n\big|_{k_1=1/\kappa}=0, \eqno(A.28)$$
$${\partial T_n\over\partial k_1}\big|_{k_1=1/\kappa}=0, \eqno(A.29)$$
$$ T_n\big|_{k_1=k_2}=-12k_1^2(1-\kappa^2k_1^2)^5
\prod_{i=3}^n(k_1^2-k_i^2)\left[(k_1^4+k_1^2k_i^2+k_i^4)\kappa^4-6(k_1^2+k_i^2)\kappa^2+9\right]\times$$
$$\times T_{n-2}(k_3, k_4, ..., k_n)=0. \eqno(A.30)$$
The symmetry and evenness of $T_n$ in $k_i\, (i=1, 2, ..., n)$ as well as the relations (A.27)-(A.30) show that $T_n$ has a factor
$$\prod_{i=1}^nk_i^2\left(k_i^2-{1\over\kappa^2}\right)^2\prod_{1\leq i<j\leq n}(k_i^2-k_j^2)^2,$$
whose degree in $k_i\ (i=1, 2, ..., n)$ is $2n^2+4n$.
On the other-hand, the degree of $T_n$ from (A.26) is $2n^2+2n+2$ at most. 
This is impossible except for $T_n\equiv 0$, completing the proof of (A.26) and hence (2.26). \hspace{\fill}$\Box$ \par
\bigskip
\leftline{\bf Acknowledgements}\par
\bigskip
This work was partially supported by JSPS KAKENHI Grant Number 22540228.  \par

\newpage
\leftline{\bf References} \par
\baselineskip=6mm
\begin{enumerate}[{[1]}]
\item Novikov V 2009 Generalizations of the Camassa-Holm equation {\it J. Phys. A: Math. Theor.} {\bf 42}  342002
\item Hone A N W and Wang J P 2008 Integrable peakon equations with cubic nonlinearity {\it J. Phys. A: Math. Theor.} {\bf 41}  372002
\item Hone A N W, Lundmark H and Szmigielski J 2009 Explicit multipeakon solutions of Novikov's cubically nonlinear integrable Camassa-Holm type equation  {\it Dynamics of PDE} {\bf 6}  253-89
\item Grayshan K 2013 Peakon solutions of the Novikov equation and properties of the data-to-solution map {\it J. Math. Anal. Appl.} {\bf 397}  515-21
\item Hirota R and Satsuma J 1976 $N$-soliton solutions of model equations for shallow water waves {\it J. Phys. Soc. Japan} {\bf 40}  611-2
\item  Matsuno Y 2005 Multisoliton solutions of the Degasperis-Procesi equation and their peakon limit {\it Inverse Problems} {\bf 21} 1553-70 
\item Matsuno Y 2005 The $N$-soliton solution of the Degasperis-Procesi equation {\it Inverse Problems} {\bf 21} 2085-101
\item  ParkerA  and  Matsuno Y 2006 The peakon limits of soliton solutions of the Camassa-Holm equation {\it J. Phys. Soc. Japan} {\bf 75} 124001
\item  Matsuno Y 2007 The peakon limit of the $N$-soliton solution of the Camassa-Holm equation  {\it J. Phys. Soc. Japn} {\bf 76} 034003 
\item Matsuno Y 1990 B\"acklund transformation, conservation laws, and inverse scattering transform of a model integrodifferential equation for water waves {\it J. Math. Phys. } {\bf 31} 2904-16
\item Fokas A 1995 The Korteweg-de Vries equation and beyond {\it Acta. Appl. Math.} {\bf 39}  295-305
\item  Fuchssteiner B 1996 Some tricks from the symmetry-toolbox for nonlinear equations: Generalizations of the Camassa-Holm equation {\it Phys. D} {\bf 95} 229-43
\item Olver P J and  Rosenau P 1996 Tri-Hamiltonian duality between solitons and solitary-wave solutions having compact support {\it Phy. Rev. E} {\bf 53} 1900-6 
\item  Qiao Z 2006 A new integrable equation with cuspons and W/M-shape-peaks solitons, {\it J. Math. Phys.} {\bf 47} 112701
\item  Qiao Z 2007 New integrable hierarchy, its parametric solutions, cuspons, one-peak solitons, and M/W-shape peak solitons, {\it J. Math. Phys.} {\bf 48} 082701
\item  Bies P M,  G\'orka P and  Reyes E G 2012 The dual modified Korteweg-de Vries-Fokas-Qiao equation: Geometry and local analysis, {\it J. Math. Phys.} {\bf 53} 073710 
\item  Ivanov R I and  Lyons T 2012 Dark solitons of the Qiao's hierarchy {\it J. Math. Phys.} {\bf 53} 123701 
\item Matsuno Y 2013 B\"acklund transformation and smooth multisoliton solutions for a modified Camassa-Holm equation with cubic nonlinearity {\it J. Math. Phys.} {\bf 54} 051504 
\item  Ablowitz M J,  Kaup D J,  Newell A  C  and  Segur H 1974 The inverse scattering transform - Fourier analysis for nonlinear problems {\it Stud. Appl. Math.} {\bf 53}, 249-315
\item Clarkson P A and Mansfields E 1994 On a shallow water wave equation {\it Nonlinearity} {\bf 7} 975-1000
\item Fisher M and Schiff J 1999 The Camassa-Holm equation: conserved quantities and the initial value problem {\it Phys. Lett.} A {\bf 259} 371-6
\item Parker A 2004 On the Camassa-Holm equation and a direct method of solution. I. Bilinear form and solitary waves {\it Proc. R. Soc. Lond.} A {\bf  460} 2929-57
\item Parker A 2005 On the Camassa-Holm equation and a direct method of solution. II. Soliton solutions {\it Proc. R. Soc. Lond.} A {\bf  461} 3611-32
\item Parker A 2005 On the Camassa-Holm equation and a direct method of solution. III. $N$-soliton solutions {\it Proc. R. Soc. Lond.} A {\bf  461} 3893-911
\item Matsuno Y 2005 Parametric representation for the multisoliton solution of the Camassa-Holm equation  {\it J. Phys. Soc. Japan} {\bf 74} 1983-7
\item Gui G L, Liu Y, Olver P J and Qu C Z  2013 Wave-breaking and peakons for a modified Camassa-Holm equation {\it Comm. Math. Phys.} {\bf 319} 731-59

\end{enumerate}
\end{document}